\documentclass[twocolumn,groupedaddress,
superscriptaddress,preprintnumbers,
amsmath,amssymb,
aps,
]{revtex4}
\usepackage{graphicx}
\usepackage{dcolumn}
\usepackage{bm}%
\usepackage[colorlinks=true,citecolor=blue]{hyperref}
\hypersetup{colorlinks=true,citecolor=blue,linkcolor=red,urlcolor=blue}

\usepackage{float}
\usepackage{amsmath}

\usepackage{color}
\definecolor{Green}{RGB}{0,204,102}
\definecolor{Purple}{RGB}{102,0,255}
\definecolor{Blue}{RGB}{51,153,255}
\definecolor{Red}{RGB}{151,010,010}

\makeatletter
\def\@bibdataout@aps{%
\immediate\write\@bibdataout{%
@CONTROL{%
apsrev41Control%
\longbibliography@sw{%
    ,author="08",editor="1",pages="1",title="0",year="1"%
    }{%
    ,author="08",editor="1",pages="1",title="",year="1"%
    }%
  }%
}%
\if@filesw \immediate \write \@auxout {\string \citation {apsrev41Control}}\fi 
}
\makeatother

\begin{document}

\title{Strain and electric-field control of spin-spin interactions in monolayer CrI$_3$}

\author{Sahar Izadi Vishkayi}
\affiliation{School of Nano Science, Institute for Research in Fundamental Sciences (IPM), Tehran 19395-5531, Iran}
\author{Zahra Torbatian}
\affiliation{School of Nano Science, Institute for Research in Fundamental Sciences (IPM), Tehran 19395-5531, Iran}
\author{Alireza Qaiumzadeh}
\affiliation{Center for Quantum Spintronics, Department of Physics, Norwegian University of Science and Technology, NO-7491 Trondheim, Norway}
\author{Reza Asgari}
\email{asgari@ipm.ir}
\affiliation{School of Nano Science, Institute for Research in Fundamental Sciences (IPM), Tehran 19395-5531, Iran}
\affiliation{School of Physics, Institute for Research in Fundamental Sciences (IPM), Tehran 19395-5531, Iran}
\affiliation{ARC Centre of Excellence in Future Low-Energy Electronics Technologies, UNSW Node, Sydney 2052, Australia}

\begin{abstract}
We investigate the impact of mechanical strains and a perpendicular electric field on the electronic and magnetic ground-state properties of two-dimensional monolayer CrI$_3$ using density functional theory. We propose a minimal spin model Hamiltonian, consisting of symmetric isotropic exchange interactions, magnetic anisotropy energy, and Dzyaloshinskii-Moriya (DM) interactions, to capture most pertinent magnetic properties of the system.
We compute the mechanical strain and electric field dependence of various spin-spin interactions.
Our results show that both the amplitudes and signs of the exchange interactions can be engineered by means of strain, while the electric field affects only their amplitudes. However, strain and electric fields affect both the directions and amplitudes of the DM vectors. The amplitude of the magnetic anisotropy energy can also be substantially modified by an applied strain.
We show that in comparison with an electric field, strain can be more efficiently used to manipulate the magnetic and electronic properties of the system. Notably, such systematic tuning of the spin interactions is essential for the engineering of room-temperature spintronic nanodevices.

\end{abstract}

\maketitle


\section{Introduction}\label{sec:intro}
In the past decade, the field of two-dimensional (2D) crystalline materials has seen rapid and almost revolutionary development \cite{ref1}. Undoubtedly, the success in this field owes much to ground-breaking advances in experimental techniques. Surprisingly, Huang et al. \cite{huang2017layer} demonstrated monolayer chromium triiodide (CrI$_3$) as an Ising-like 2D hexagonal ferromagnetic crystal, showing the removal of a restriction of the Mermin-Wagner theorem \cite{ref2} in the CrI$_3$ crystal. They observed that the magnetic order of the crystal is a layer-dependent phenomenon and recognized the presence of a large magnetocrystalline anisotropy, which effectively lifts the invariance under rotations.

The combination of the magnetic and other unique properties of 2D materials has rapidly attracted the attention of researchers to novel magnetic 2D materials \cite{2Dreview1,2Dreview2,2Dreview3,2Dreview4,burch2018magnetism, gong2019two}. In this context, CrI$_3$ exhibits a plethora of intriguing properties \cite{huang2017layer,klein2018probing,wang2018very,huang2018electrical, wang2016doping,sivadas2018stacking,PhysRevMaterials.3.031001,PhysRevB.98.104307,kim2018one}. For instance, the magnetization of monolayer CrI$_3$ is remarkably saturated by doping \cite{ref4}. In a bilayer CrI$_3$ system, on the other hand, the interlayer magnetic order significantly depends on the doping type.

The control of spin-spin interactions in magnetic systems is an essential topic related to the fundamental physics of quantum magnetism as well as applied spintronics-based technology. Different magnetic phases and exotic spin textures may be realized through the engineering of spin interactions in low-dimensional magnetic materials.

Recently, it has become evident that the magnetic properties of 2D van der Waals heterostructures can be controlled by applying an external electric field \cite{huang2018electrical, ref5, ref10}. This is an important capability in spintronic and logic/memory devices
\cite{huang2018electrical}.
The magnetic properties of monolayer \cite{ref12, ref13, ref14} and bilayer \cite{ref11} CrI$_3$ have been studied by various research groups in order to classify their magnetic orders using density functional theory (DFT). These studies have found that an external electric field can drive a transition from an antiferromagnetic (AFM) to a ferromagnetic (FM) phase in a bilayer system \cite{ref5, huang2018electrical,ref11}. The effect of a perpendicular electric field on the nearest-neighbor Dzyaloshinskii-Moriya (DM) interaction has also been studied through ab initio calculations \cite{Liu2018,Behera,Ghosh}; however, there are noticeable discrepancies among the reported results. Furthermore, the presence of an {\it{intrinsic}} out-of-plane DM interaction has not been reported in those studies, although such an interaction has been predicted to exist in hexagonal lattices on the basis of microscopic calculations \cite{Alireza} and symmetry arguments \cite{owerre2016first, Yaroslav}.

Making use of an external mechanical strain is another efficient method of controlling the electronic and magnetic properties of 2D materials. The magnetic properties of monolayer and bilayer CrI$_3$ depend on the applied strain \cite{leon2020strain}. Recently, an FM-AFM transition \cite{LucaWebprb, liu2018multi, Wu2019} and a decrease in the energy bandgap \cite{LucaWebprb, Wu2019} under the exertion of an external strain have been reported. The strain dependence of the CrI$_3$ phonon spectra has also been studied \cite{larson2018raman}. However, no systematic study has been presented on the effects of uniaxial and biaxial strains on the extrinsic and intrinsic DM interactions in monolayer CrI$_3$.

In this paper, we present a comprehensive study on the electronic and magnetic properties of monolayer CrI$_3$, as a representative of 2D transition metal trihalides, in the presence of mechanical strain and electric fields by means of DFT calculations. We invoke the DFT results to extract a suitable spin model Hamiltonian of CrI$_3$ that reproduces accurate and viable magnetic properties of the system. Having calculated the band structures of monolayer CrI$_3$ under exposure to external electric fields and strains, we then compute the isotropic and anisotropic symmetric exchange interactions, DM interactions, and anisotropy energy.

This paper is organized as follows. We commence with a description of our theoretical formalism in Sec. II, followed by the details of the DFT simulations and spin model Hamiltonian. Numerical results for the band structures and spin-spin interaction parameters in the presence of electric fields and strains are reported in Sec. III. We summarize our main findings in Sec. IV.

\section{Theoretical and Computational Methods}\label{sec:theory}

\begin{figure}
\begin{center}
\includegraphics*[width=8.cm]{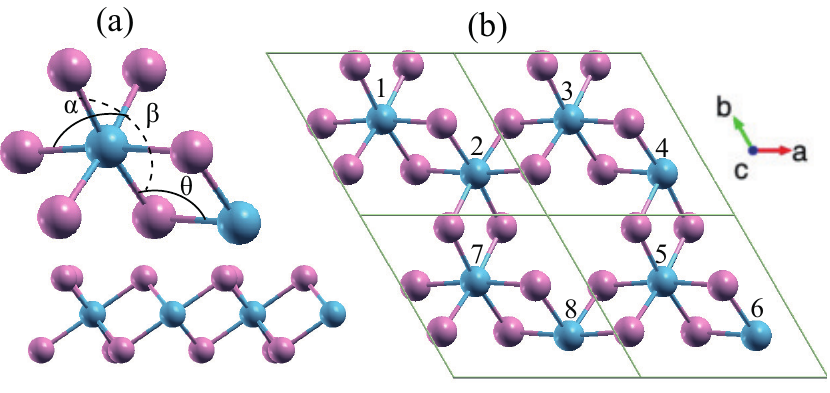}
\caption{(Color online) (a) Top and side views of monolayer CrI$_3$. The bonding angles between the atoms in the monolayer are denoted by $\theta$, $\alpha$ and $\beta$. (b) A $2\times2\times1$ supercell of the monolayer. The Cr atoms are numbered as shown in Table \ref{sforJ}. The blue (pink) spheres represent Cr (I) atoms. }
\label{geometryCrI3}
\end{center}
\end{figure}

A monolayer of CrI$_3$ in the $x-y$ plane, consisting of three atomic layers, is considered, as illustrated in Fig. \ref{geometryCrI3}(a). The hexagonal unit cell comprises six iodine atoms and two chromium atoms, where the iodine atoms are attached to the chromium atoms in accordance with the octet rule. Our analyses are based on DFT calculations performed using the Quantum Espresso package \cite {refQE}, in which norm-conserving pseudopotentials are used to determine the electron-ion interactions. We use the Perdew-Burke-Ernzerhof (PBE) functional \cite{refpbe} as the generalized gradient exchange-correlation approximation and a plane-wave cutoff energy of $80$ Ry. To avoid any interactions between the plane images, a 25 \AA\, vacuum is applied along the $z$-axis. To calculate the ground-state energy, an $8\times8\times1$ $k$-point mesh grid is used within the first Brillouin zone.
To obtain a reliable total ground-state energy, we maintain a high degree of accuracy of $10^{-10}$ eV. Furthermore, the unit cell and atomic positions are optimized until the maximum force on each atom becomes less than $10^{-3}$ eV/\AA.
The total energy is computed by means of fully relativistic self-consistent-field DFT calculations incorporating the spin-orbit coupling (SOC) and non-collinear spin-polarization effects in order to obtain the magnetic anisotropy energy (MAE) and DM interactions.

\begin{figure}
\begin{center}
\includegraphics*[width=7.8cm]{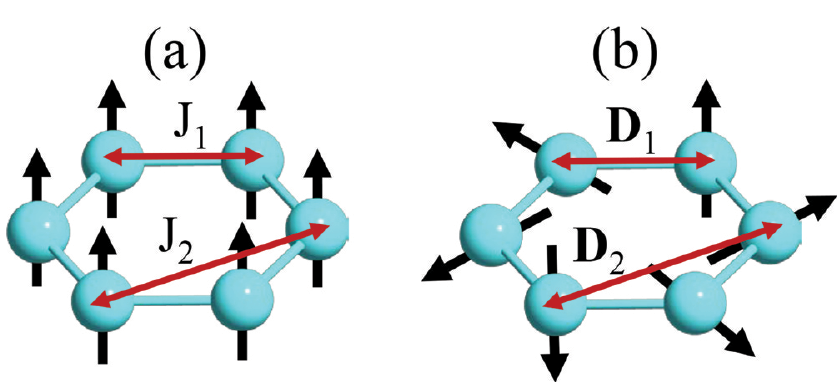}
\caption{(Color online) (a) Schematic picture of the symmetric exchange couplings between Cr atoms, where $J_1$ denotes the coupling between nearest-neighbor atoms and $J_2$ denotes the coupling between next-nearest-neighbor atoms. (b) The same as (a) for the DM vectors.}
\label{geometrycoupling}
\end{center}
\end{figure}

To compute the spin-spin interactions in a 2D CrI$_3$ crystal through ab initio calculations, we use a minimal spin Hamiltonian for a 2D magnetic hexagonal lattice \cite{Alireza, Hxxz, LebingChen}:

\begin{equation}
{\cal H}=\sum_{i,j} \left( \frac{1}{2}J_{ij}{\bf S}_i\cdot{\bf S}_j+  {\gamma}_{ij}S_{iz}S_{jz}+ \frac{1}{2}{\bf D}_{ij}\cdot({\bf S}_i\times{\bf S}_j )\right),
 \label{eq1}
\end{equation}
where ${\bf S}_i$ denotes the spin of the $i^{\text {th}}$ Cr atom, $J_{ij}$ is the symmetric Heisenberg exchange coupling between atoms $i$ and $j$, ${\gamma}_{ij}$ is the anisotropy energy along the $z$-axis, and ${\bf D}_{ij}$ is the DM vector. The direction of the DM vector is dictated by the symmetry of the magnetic crystal. The magnetic coupling parameters are shown in Fig.~\ref{geometrycoupling}.

It has been shown that in a hexagonal lattice structure and in the presence of inversion symmetry, a finite intrinsic DM vector perpendicular to the plane arises from the next-nearest-neighbor intrinsic SOC \cite{Alireza, Yaroslav, owerre2016first}. This particular intrinsic DM interaction in hexagonal lattices leads to a few fascinating topological properties, such as topological magnon insulators \cite{LebingChen, elyasi2019topologically, owerre2016first,Yaroslav}, the magnon spin Nernst effect \cite{cheng2016spin, zyuzin2016magnon}, and chiral phonon transport \cite{thingstad2019chiral}.
An extrinsic nearest-neighbor DM interaction can also be induced in this system by breaking the inversion symmetry. In this case, the DM vector lies within the plane.
The magnetic ground state, the magnetic phase transitions, and the existence of exotic magnetic textures are governed by the relative signs and ratios between the competing DM, exchange, and anisotropy interactions. As a result, it is critically important to find optimal methods for controlling the spin interactions in 2D hexagonal magnetic systems.

In the rest of this section, using the spin Hamiltonian of Eq. \ref{eq1}, we derive the necessary equations to find the spin interactions. In the next section, we use the obtained equations to numerically compute the spin interactions in CrI$_3$ by invoking numerical DFT results.

\begin{table}
\caption{Four different spin configurations (Cr1-Cr4) of the eight Cr atoms in the supercell for evaluating exchange interactions. $\uparrow$ ($\downarrow$) represents that the spin orientation of the Cr atom is parallel (antiparallel) to the $z$-direction.}
\begin{center}
\begin{tabular}{ccccccccc}
\hline
\hline
Cr & \hspace{0.2cm}1 \hspace{0.2cm}& \hspace{0.2cm}2\hspace{0.2cm} & \hspace{0.2cm}3\hspace{0.2cm} & \hspace{0.2cm}4\hspace{0.2cm} & \hspace{0.2cm}5\hspace{0.2cm} & \hspace{0.2cm}6\hspace{0.2cm} & \hspace{0.2cm}7\hspace{0.2cm} & \hspace{0.2cm}8\hspace{0.2cm} \\
\hline
Cr1 & $\uparrow$   & $\uparrow$   & $\uparrow$   & $\uparrow$   & $\uparrow$   & $\uparrow$   & $\uparrow$ & $\uparrow$ \\
\hline
Cr2 & $\uparrow$   & $\downarrow$ & $\uparrow$   & $\downarrow$ & $\uparrow$   & $\downarrow$ & $\uparrow$ & $\downarrow$ \\
\hline
Cr3 & $\uparrow$   & $\uparrow$   & $\downarrow$ & $\downarrow$ & $\downarrow$& $\downarrow$ & $\uparrow$   & $\uparrow$   \\
\hline
Cr4 & $\downarrow$ & $\uparrow$   & $\downarrow$ & $\uparrow$   & $\downarrow$& $\uparrow$ & $\uparrow$   & $\uparrow$    \\
\hline
\hline
\end{tabular}
\end{center}
\label{sforJ}
\end{table}

To calculate the symmetric exchange interactions for nearest neighbors, $J_1$, and next-nearest neighbors, $J_2$, we need a $2\times2\times1$ supercell, as shown in Fig. \ref{geometryCrI3}(b), with four different spin configurations per eight Cr atoms. The total energies of the considered spin configurations, shown in Table \ref{sforJ}, are given by

\begin{equation}
 \begin{gathered}
E_{\mathrm{Cr1}}=\frac{1}{2}[(+3\times8) J_1 S^2+(+6\times8) J_2 S^2]+8{\gamma}S^2+E_0,\\
E_{\mathrm{Cr2}}=\frac{1}{2}[(-3\times8) J_1 S^2+(+6\times8) J_2 S^2]+8{\gamma}S^2+E_0,\\
E_{\mathrm{Cr3}}=\frac{1}{2}[(+1\times8) J_1 S^2+(-2\times8) J_2 S^2]+8{\gamma}S^2+E_0,\\
E_{\mathrm{Cr4}}=\frac{1}{2}[(-3\times4) J_1 S^2+(+3\times8) J_2 S^2]+8{\gamma}S^2+E_0,
 \end{gathered}
 \label{eqEs}
\end{equation}
where $S=3/2$ is the spin of a Cr atom and $E_0$ is the nonmagnetic constant part of the energy. Making use of the mapping between the total energies obtained from DFT calculations and the spin model Hamiltonian for different states, $J_1$ and $J_2$ are eventually calculated as follows:
\begin{align}
J_1&=\frac{E_{\mathrm{Cr1}}-E_{\mathrm{Cr2}}}{24S^2},\\
J_2&=\frac{(E_{\mathrm{Cr1}}+E_{\mathrm{Cr2}} )-(E_{\mathrm{Cr3}}+E_{\mathrm{Cr4}})}{44S^2}-\frac{E_{\mathrm{Cr1}}-E_{\mathrm{Cr2}}}{(12\times44)S^2}.
\end{align}

The MAE is obtained by computing the energy difference between two perpendicular magnetic phases of the Cr atoms in the unit cell. The total energies of the out-of-plane ($E_{\perp}$) and in-plane ($E_\parallel$) phases are expressed as
\begin{align}
E_{\perp}&=\frac{1}{2}[(2\times3) J_1 S^2]+2{\gamma}S^2+E_0,\label{eq2}\\
E_{\parallel}&=\frac{1}{2}[(2\times3) J_1 S^2]+E_0,\label{eq3}
\end{align}
and the MAE is defined as $\mathrm{MAE}=E_\perp-E_\parallel$. We define the effective out-of-plane anisotropy coefficient as $\gamma=\mathrm{MAE}/2 S^2$. We follow the method presented in Ref. \cite{refMAE} to compute the MAE, where Andersen's local force theorem~\cite{andersen19805}, which is implemented in the Quantum Espresso package, is applied in two steps: i) a self-consistent calculation without the SOC is carried out to find the charge density and the spin-moment distribution, and ii) the spin moments of the Cr atoms are rotated to a certain direction, and non-self-consistent calculations are performed with the SOC term. In this step, the band energies are calculated for the in-plane and out-of-plane spin directions; thus, the difference in the band energies between the two spin moment directions provides us with the MAE.

To compute the DM interactions between nearest neighbors, ${\bf D}_1$, and next-nearest neighbors, ${\bf D}_2$, we need to consider at least a $2\times1\times1$ supercell. In the considered supercell, the Cr atoms are labeled with numbers 1-4, as shown in Fig. \ref{geometryCrI3} (b). We use the spin Hamiltonian of Eq. \ref{eq1} to find the total energy of the supercell as

\begin{align}
E= &E_0+J_1 A_{J_1}+ J_2 A_{J_2} +\gamma A_\gamma +\nonumber\\
&({\bf D}_1\cdot{\bf A}_{D_1})+({\bf D}_2\cdot{\bf A}_{D_2}),
\label{eq4}
\end{align}
where we define the following coefficients:
\begin{align}
A_{J_1}&= 2 {\bf S}_1\cdot{\bf S}_2+ {\bf S}_1\cdot{\bf S}_4 + {\bf S}_2\cdot {\bf S}_3+ 2{\bf S}_3\cdot{\bf S}_4,\nonumber\\
A_{J_2}&= 4 {\bf S}_1\cdot{\bf S}_3+ 4{\bf S}_2\cdot{\bf S}_4+ S_1^2+ S_2^2+ {S_3}^2+ {S_4}^2,\nonumber\\
A_\gamma &=4S_{1z}S_{1z}+4S_{2z}S_{2z}+4S_{3z}S_{3z}+4S_{4z}S_{4z},\nonumber\\
{\bf A}_{D_1}&=2 {\bf S}_1\times{\bf S}_2 + {\bf S}_1\times{\bf S}_4+{\bf S}_2\times{\bf S}_3+2{\bf S}_3\times{\bf S_4},\nonumber\\
{\bf A}_{D_2}&=4{\bf S} _1\times{\bf S}_3+4{\bf S}_2\times{\bf S}_4.
 \label{eq5}
\end{align}

\begin{table}[H]
\caption{The six different spin configurations of the four Cr atoms in the supercell for evaluating ${\bf D}_1$ and ${\bf D}_2$. The columns present the polar ($\vartheta$) and azimuthal ($\phi$) angles of the spin moment $S_i$ of the $i^{\text th}$  Cr atom.}
\begin{center}
\begin{tabular}{cccccccc}
\hline
\hline
 & &         &          & configurations      &       &  &   \\
 & & $(\vartheta_1, \phi_1)$& $(\vartheta_2, \phi_2)$& $(\vartheta_3, \phi_3)$& $(\vartheta_4, \phi_4)$& $(\vartheta_5, \phi_5)$& $(\vartheta_6, \phi_6)$ \\
\hline
  & {$S_1$} & $(\frac{\pi}{2}, \frac{\pi}{6})$ &  $(\frac{\pi}{2}, -\frac{\pi}{6})$ &   $(0 ,\frac{\pi}{2} )$       &   $(0 ,\frac{\pi}{2} )$      &   $(0, 0)$     & $(0,0)$   \\
 ${\bf D}_1$ & {$S_2$}  & $(\frac{\pi}{2}, 0)$ &  $(\frac{\pi}{2}, 0 )$ &    $(\frac{\pi}{6},\frac{\pi}{2} )$       &   $(-\frac{\pi}{6} ,\frac{\pi}{2} )$      &   $(\frac{\pi}{6}, 0)$     & $(-\frac{\pi}{6},0)$   \\
&{$S_3$}  & $(\frac{\pi}{2}, \frac{\pi}{6})$ &  $(\frac{\pi}{2}, -\frac{\pi}{6})$ &   $(0 ,\frac{\pi}{2} )$       &   $(0 ,\frac{\pi}{2} )$      &   $(0, 0)$     & $(0,0)$   \\
&{$S_4$}  & $(\frac{\pi}{2}, 0)$ &  $(\frac{\pi}{2}, 0 )$ &    $(\frac{\pi}{6},\frac{\pi}{2} )$       &   $(-\frac{\pi}{6} ,\frac{\pi}{2} )$      &   $(\frac{\pi}{6}, 0)$     & $(-\frac{\pi}{6},0)$   \\

\hline

&{$S_1$} & $(\frac{\pi}{2}, \frac{\pi}{6})$ &  $(\frac{\pi}{2}, -\frac{\pi}{6})$ &   $(0 ,\frac{\pi}{2} )$       &   $(0 ,\frac{\pi}{2} )$      &   $(0, 0)$     & $(0,0)$   \\
${\bf D}_2$& {$S_2$} & $(\frac{\pi}{2}, \frac{\pi}{6})$ &  $(\frac{\pi}{2}, -\frac{\pi}{6})$ &   $(0 ,\frac{\pi}{2} )$       &   $(0 ,\frac{\pi}{2} )$      &   $(0, 0)$     & $(0,0)$   \\
&{$S_3$}  & $(\frac{\pi}{2}, 0)$ &  $(\frac{\pi}{2}, 0 )$ &    $(\frac{\pi}{6},\frac{\pi}{2} )$       &   $(-\frac{\pi}{6} ,\frac{\pi}{2} )$      &   $(\frac{\pi}{6}, 0)$     & $(-\frac{\pi}{6},0)$   \\
&{$S_4$}  & $(\frac{\pi}{2}, 0)$ &  $(\frac{\pi}{2}, 0 )$ &    $(\frac{\pi}{6},\frac{\pi}{2} )$       &   $(-\frac{\pi}{6} ,\frac{\pi}{2} )$      &   $(\frac{\pi}{6}, 0)$     & $(-\frac{\pi}{6},0)$   \\

\hline
\hline
\end{tabular}
\end{center}
\label{spinforD1}
\end{table}

${\bf D}_1$ is read out by considering six different spin configurations of the Cr atoms in the supercell, as shown in the first four rows of Table \ref{spinforD1}. The total energy of each configuration can be obtained through DFT by applying the SOC. Once the coefficients have been calculated from Eq. \ref{eq5}, $A_{J1}$, $A_{J2}$ and $A_\gamma$ can be evaluated for all considered configurations. ${\bf A}_{D2}$ is zero because the next-nearest neighbors' spins are parallel; thus, the nearest-neighbor DM interaction is obtained. Finally, we can find ${\bf D}_2$ by considering the magnetization moments of the Cr atoms in the supercell in accordance with the last four rows of Table \ref{spinforD1}, where the spin moments of the nearest neighbors are parallel.

\section{\label{sec:level3} Numerical Results and Discussions}
In this section, we first present the magnetic ground state, electronic band structure, and atomic orbital characteristics of a free monolayer of CrI$_3$. Next, we present the effects of uniaxial and biaxial mechanical strains on the spin interactions. Finally, we discuss the renormalization of the spin interactions in the presence of a perpendicular electric field.

To explore the electronic and magnetic ground states of the system, we consider two different magnetic phases, namely, the FM and AFM states, in which the spin moments of the Cr atoms are aligned in parallel and antiparallel directions, respectively. Spin-dependent DFT calculations show that monolayer CrI$_3$ is well built in the FM phase compared with the AFM phase, with a total energy difference of $22.36$ meV per Cr atom. Our numerical results show that the lattice vectors of the unit cells are not significantly different between the FM and AFM phases, with the change in the bonding length being approximately $0.04 \%$.
\begin{figure}
 \includegraphics{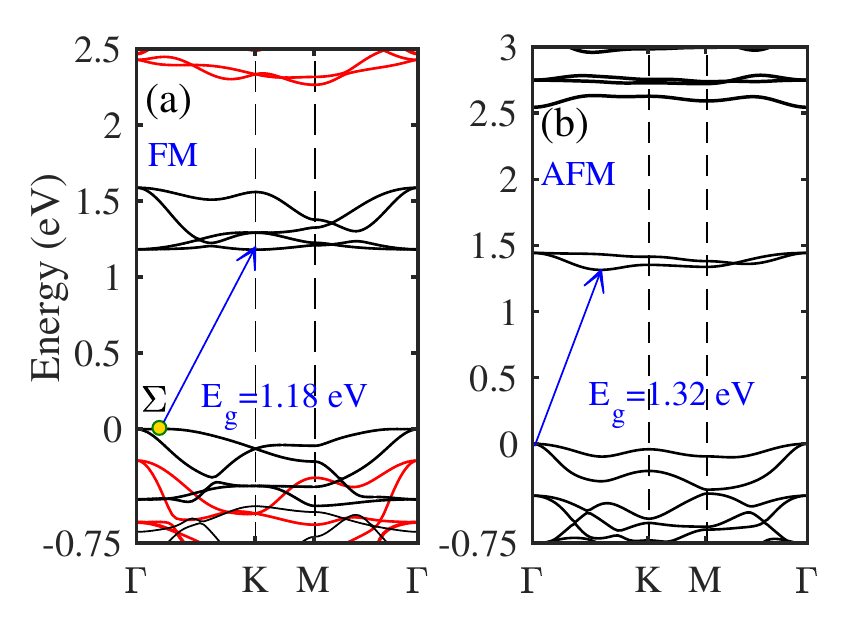}
\caption{(Color online) The band structures of monolayer CrI$_3$ along the high-symmetry $k$-point of the first Brillouin zone in (a) the half-semiconductor FM configuration and (b) the semiconductor AFM configuration with an indirect bandgap. The black (red) lines represent spin-up (spin-down) energy bands in (a). The VBM is shifted to zero energy. Spin-dependent DFT calculations show that the FM configuration has a lower ground-state energy than the AFM configuration in monolayer CrI$_3$. The VBM is connected to the CBM by an arrow.} \label{fig:2}
 \label{bsforFMandAFM}
\end{figure}

Figures \ref{bsforFMandAFM}(a) and (b) show the band structures of the monolayer in the FM and AFM phases, respectively. It is clear that FM monolayer CrI$_3$ is a half-semiconductor with an indirect bandgap, as demonstrated by the arrow in Fig. \ref{bsforFMandAFM}(a). For the spin-up bands, with a bandgap of $E_g=1.18$ eV, the valence band maximum (VBM) is located at the $\Sigma$ point along the $\Gamma$-K path of the first Brillouin zone, and the conduction band minimum (CBM) is located at the K point. For the spin-down bands, with $E_g=2.47$ eV, the VBM is located at the $\Gamma$ point, while the CBM is at the M point. These results are in good agreement with those obtained in \cite{refgaptheory}. Figure \ref{bsforFMandAFM}(b) shows that the AFM phase of CrI$_3$ is an indirect-bandgap semiconductor with a bandgap of 1.32 eV and that its spin-degenerate CBM is at the $\Gamma$ point, while the VBM is located between the $\Gamma$ and K points.

\begin{figure}[!t]
\begin{center}
 \includegraphics{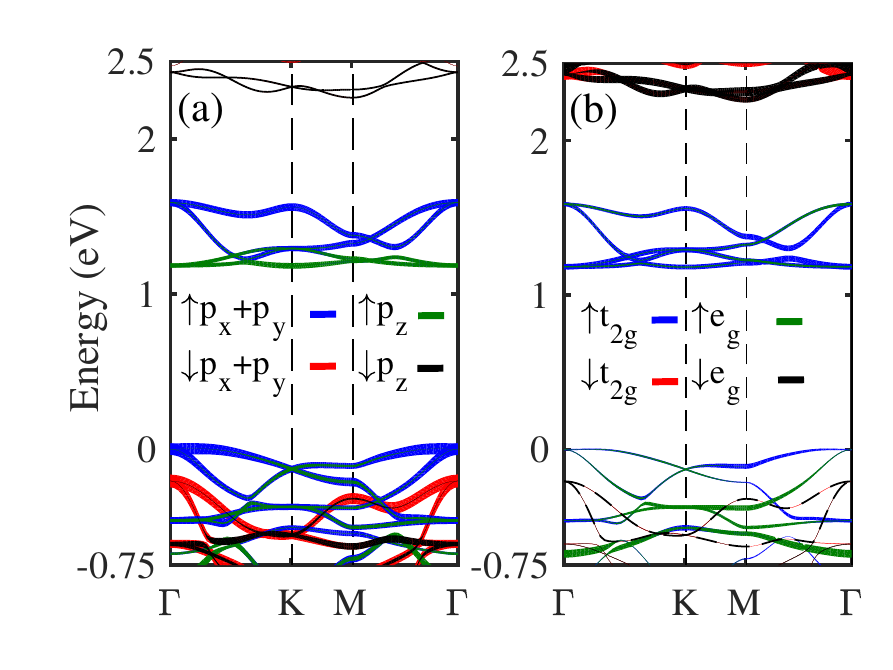}
\caption{ (Color online) The atomic orbital characteristics illustrated with respect to the band structure. The band structure is projected over (a) the $p_z$ and $p_x+p_y$ orbitals of the iodine atoms and (b) the $\mathrm{t_{2g}}$ and $\mathrm{e_g}$ orbitals of the chromium atoms in the FM configuration of monolayer CrI$_3$.}
 \label{bsforCr_I}
\end{center}
\end{figure}

\begin{table}
\caption{(Color online) The contributions of the $p$ orbitals of the iodine atoms and the $d$ orbitals of the chromium atoms (in percent) to the VBMs and CBMs of the spin-up and spin-down bands of FM monolayer CrI$_3$.}
\begin{center}
\begin{tabular}{cccccccccc}
  \hline
  \hline

   state        &   $k$-point &~~$p_z~$   &~~$p_x~$    &~~$p_y~$   & $~~d_{z^2}~$   & $d_{x^2-y^2}$   & $~~d_{xy}$ &~~$d_{zx}$ &~~$d_{zy}$  \\
\hline
  VBM$\uparrow$ &   $\Sigma$       &  0        &   64       &  23       &    0           &     3           &    4       &     2     &  4    \\
  \hline
  CBM$\uparrow$ &  $K$    &  45 &   2        &   2       &    0           &   10            &   10       &  14       &   14   \\
  \hline
  VBM$\downarrow$&   $\Gamma$ &   0       &  43         &  54      &     0          &  1              &    1       &    0      &  0 \\
  \hline
  CBM$\downarrow$ &  M     &  2        &  1          &   5      &     40         &  9              &   0        &    42     &  0  \\

  \hline
  \hline
\end{tabular}
\end{center}
\label{tabelof_orb_decompose}
\end{table}

Now, we explore the atomic orbital characteristics in relation to the band structure of the system. The VBM and CBM are mainly formed by the hybridization of the $p$ orbitals of the iodine atoms and the $d$ orbitals of the chromium atoms (see Table \ref{tabelof_orb_decompose} for more details). In addition, Figs. \ref{bsforCr_I}(a) and (b) show the participation of the $p$ orbitals of the iodine atoms and the $d$ orbitals of the chromium atoms, respectively. Notably, the spin-up conduction band is predominantly composed of the $p_z$ orbitals of the iodine atoms, especially at the CBM, while the $p_x$ and $p_y$ orbitals generate the VBM. The octahedral environment around the chromium atoms creates a strong crystal field that splits the $3d$ orbitals of Cr into $\mathrm{e_g}$ orbitals ($d_{z^2}$ and $d_{x^2-y^2}$) and $\mathrm{t_{2g}}$ orbitals ($d_{zx}$, $d_{zy}$, and $d_{xy}$). Furthermore, the $3d$ orbitals of the chromium atoms represent the most important contribution to the conduction bands. Figure \ref{bsforCr_I}(b) shows that the role of the $\mathrm{e_g}$ orbitals of the Cr atoms in the valence and conduction bands is much smaller than that of the $\mathrm{t_{2g}}$ orbitals. The CBM mainly originates from the $\mathrm{t_{2g}}$ orbitals of the Cr atoms and the $p_z$ orbitals of the I atoms, while the valence band is formed through the hybridization of the $p_x+p_y$ orbitals of the iodine atoms and the $d$ orbitals of the chromium atoms.
According to Hund’s rule, the $\mathrm{t_{2g}}$ orbitals are occupied by 3 electrons, and $S=3/2$ for the Cr atoms.
The results show that the spin-down valence bands originate from the $p$ orbitals of the iodine atoms, while the spin-down conduction bands arise from the $3d$ orbitals of the Cr atoms.

\subsection{Effects of biaxial strains on the electronic and magnetic properties}\label{sec:level3}

\begin{figure}
\begin{center}
 \includegraphics*[width=8.4cm]{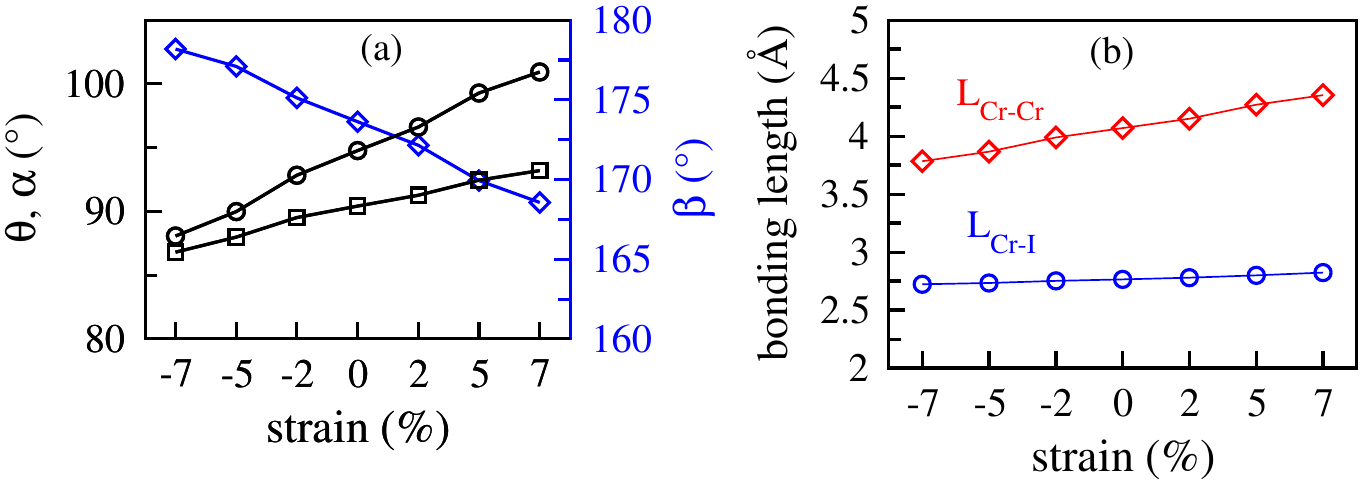}
\caption{(Color online) (a) The bonding angles ($\theta$ -- circles, $\alpha$ -- squares, and $\beta$ -- diamonds) and (b) the chromium-iodine bonding length (L$_{\text Cr-I}$) and chromium-chromium distance (L$_{\text Cr-Cr}$) in monolayer CrI$_3$ under compressive (negative sign) and tensile (positive sign) biaxial strains. The $\beta$ angle increases toward a straight angle with increasing compressive strain, while it decreases with increasing tensile strain. The $\alpha$ and $\theta$ angles, on the other hand, decrease with increasing compressive strain and increase with increasing tensile strain. Note that the Cr-Cr distance decreases under compressive strain. }\label{fig:4}
 \label{LStrain}
\end{center}
\end{figure}

In this subsection, we investigate the effects of biaxial mechanical strains on the electronic and magnetic properties of the monolayer for both compressive and tensile strains of $2 \%$, $5 \%$, and $7 \%$.
Because of the deformations induced by these strains, the bonding angles and atomic lengths are changed, as shown in Figs. \ref{LStrain}(a) and (b). A compressive strain leads to an increase in the $\beta$ angle (Fig. \ref{geometryCrI3}(a)), whereas this angle decreases under tensile strain.
By contrast, the $\alpha$ and $\theta$ angles decrease under compressive strain and increase under tensile strain.
These effects show that strain alters the crystal structure of the monolayer and thus should affect the electronic and magnetic properties of the system.

\begin{figure}
\begin{center}
 \includegraphics*[width=8.4cm]{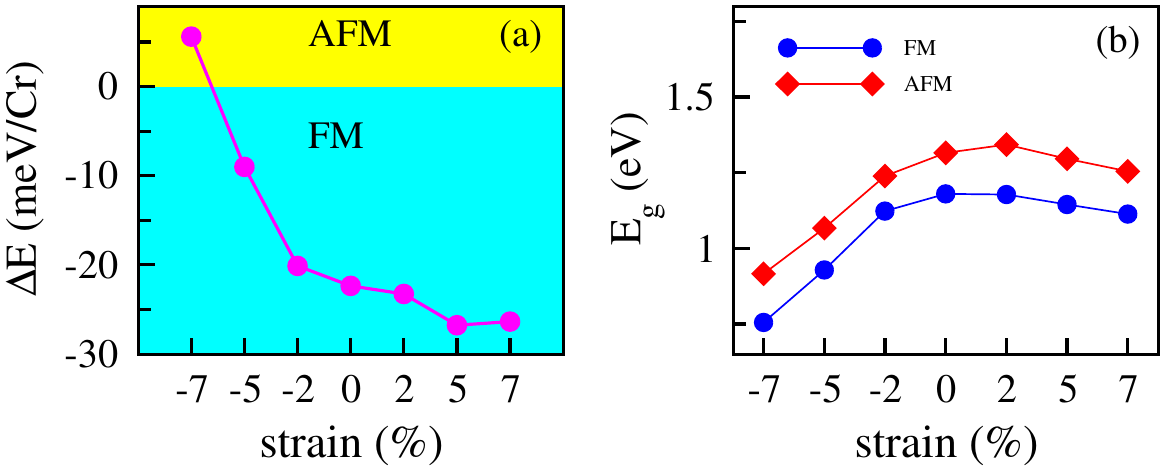}
\caption{(Color online) (a) The total energy difference between the FM and AFM configurations, $\Delta E$, and (b) the variation in the bandgap as a function of strain for monolayer CrI$_3$ in the FM and AFM configurations. Strain leads to a change in the magnetic order of the system.}\label{fig:5}
 \label{bandgapforstrain}
\end{center}
\end{figure}

To find the effect of the magnetic phase diagram of the system, we compute the total energy difference $\Delta E$ between the FM and AFM configurations in the presence of strain. The results, plotted in Fig. \ref{bandgapforstrain}(a), show a phase transition from the FM phase to the AFM phase for compressive strains greater than $7 \%$ \cite{LucaWebprb}. An alternative way to perceive the correct ground-state phase of the system is to evaluate the $\theta$ angle, as discussed in Ref. \cite{Hxxz}.

Indeed, the magnetic ground state of the system is dictated by the competition between two types of magnetic exchange interactions: first, the superexchange interactions stemming from virtual excitation through nonmagnetic ligands ($\mathrm{e_g-e_g}$), which prefer an FM alignment \cite{anderson1950}, and second, the direct magnetic exchange interactions between nearest-neighbor magnetic ions ($\mathrm{t_{2g}-t_{2g}}$), which favor an AFM configuration. Since the length of the Cr-Cr bonds decreases under compressive strain and $\theta$ deviates from 90 degrees, a transition from the FM phase to the AFM phase is probable under a sufficiently large compressive strain.

We now investigate the effects of strain on the electronic band structure of the system.
Figure \ref{bandgapforstrain}(b) shows that the bandgap rapidly (slowly) decreases under compressive (tensile) strain in both the FM and AFM configurations. In the presence of compressive strain, the bandgap decreases rapidly since the hybridization of the atomic orbitals is increased due to the reduction in the bonding length.

\begin{figure}
\begin{center}
 \includegraphics*[width=9cm]{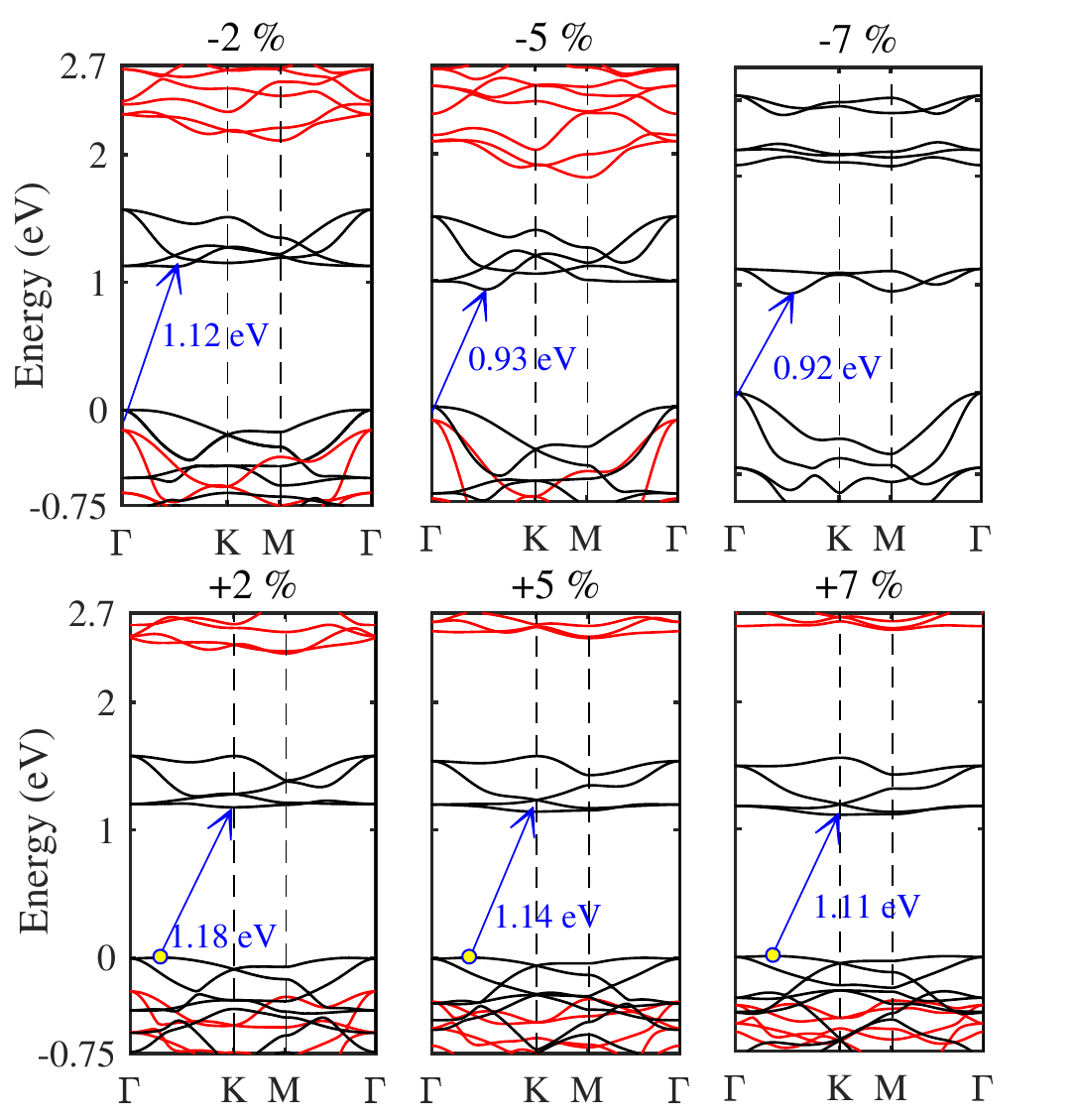}
\caption{(Color online) The band structures of monolayer CrI$_3$ under $2 \%$, $5 \%$ and $7 \%$ compressive (upper panels) and tensile (lower panels) biaxial strains. The black (red) lines represent spin-up (spin-down) bands, and the VBM is connected to the CBM by an arrow. The FM phase is considered for all strains except the $7 \%$ compressive strain, for which the ground state is AFM.}
 \label{bsforstrainFM}
\end{center}
\end{figure}

Figure \ref{bsforstrainFM} shows that a compressive strain causes the CBM and VBM of the spin-up bands within the first Brillouin zone to shift; however, the monolayer remains an indirect-bandgap semiconductor within the range of applied strains considered here.
The results reveal that at the CBM of the spin-up bands, the contribution of the iodine $p_z$ orbitals decreases, while the participation of the $p_x+p_y$ orbitals increases.

Figure \ref{bsforCr_I}(b) shows that the contribution of the $d$ orbitals of the Cr atoms (see also Table \ref{tabelof_orb_decompose}) in the pristine layer of CrI$_3$ causes the curvature of the spin-up CBM to decrease. Therefore, the spin-up electrons in these flat bands have a larger effective mass and are almost localized in the free layer. Our calculations show that a compressive strain reduces the contribution of the $d$ orbitals of the Cr atoms and thus increases the contribution of the iodine $p_x+p_y$ orbitals at the spin-up CBM. Consequently, the band curvature is increased, and the electron localization is reduced. The increase in the band curvature of the spin-up electrons close to the CBM in the presence of a compressive strain is vividly illustrated in Fig. \ref{bsforstrainFM}.

The contribution of the delocalized $p_x+p_y$ orbitals of the Cr atoms at the spin-up VBM in the pristine monolayer is given in Table \ref{tabelof_orb_decompose}. Our calculations show that a compressive strain increases the contribution of the $p_x+p_y$ orbitals while decreasing the partial contribution of the localized $d$ orbitals. Therefore, the band curvature near the VBM is increased by a compressive strain, as shown in Fig. \ref{bsforstrainFM}.

In contrast to the monolayer's response to a compressive strain, the spin-up conduction bands remain flat under the application of a tensile strain.
However, a tensile strain increases the contribution of the $d$ orbitals at the spin-up VBM; consequently, the spin-up holes become more localized, as shown in Fig. \ref{bsforstrainFM}. The application of a tensile strain shifts the location of the spin-up VBM along the $\Gamma$-K path, while the spin-up CBM remains at the K point.

The spin-down VBM and CBM of the free layer are located at the $\Gamma$ and M points, respectively. The bandgap of the spin-down bands is reduced with the application of a compressive strain, while it is increased by a tensile strain. Therefore, the half-semiconducting behavior of monolayer CrI$_3$ is modified by a tensile strain. A tensile strain larger than $5\%$ changes the indirect bandgap of the spin-down bands of the free monolayer (2.47 eV) into a direct bandgap (2.83 eV) at the M point.

For completeness in our discussion of the effects of strain on monolayer CrI$_3$, we should emphasize that the ground state of the free layer is an FM state.
In the AFM configuration, the system remains an indirect-bandgap semiconductor in the presence of both compressive and tensile
strains of less than $7 \%$. However, the monolayer becomes a direct-bandgap semiconductor, with both the CBM and VBM located at the K point, under the application of a tensile strain of greater than $7 \%$.

\begin{figure}
\begin{center}
\includegraphics*[width=8.4cm]{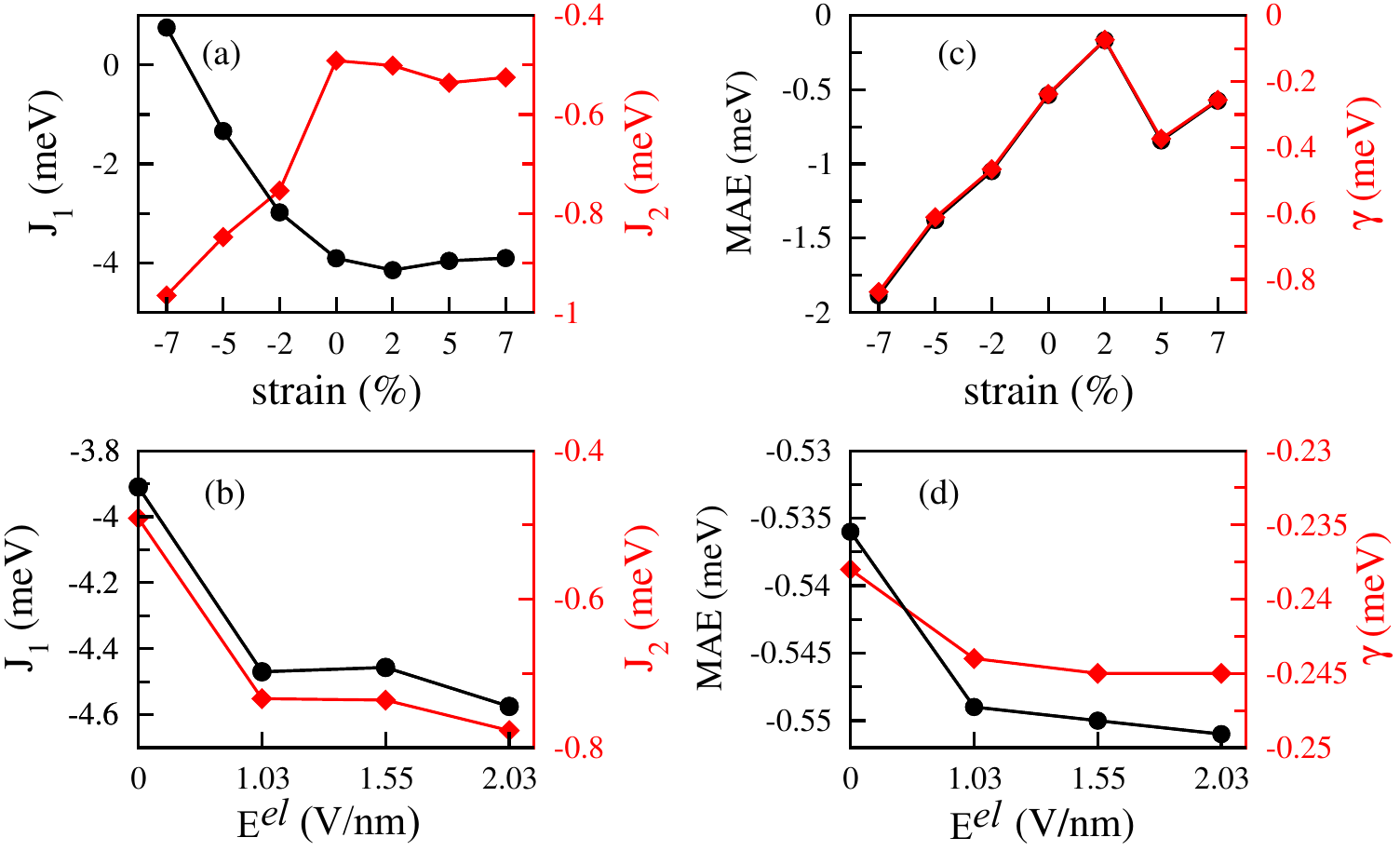}
\caption{(Color online) (a) and (c) The nearest-neighbor and next-nearest-neighbors magnetic exchange couplings and (b) and (d) the MAE and magnetic anisotropy coefficients of monolayer CrI$_3$ as functions of the applied strain and electric field, respectively.}\label{fig:7}
 \label{J1graph}
\end{center}
\end{figure}

Above, we considered the effects of strain on the electronic ground state of the system. As discussed earlier, strain also affects the magnetic ground state of the system and the spin-spin interactions. Now, we compute the spin-spin interactions in the monolayer using the equations derived in the previous section. Figure \ref{J1graph}(a) shows the nearest-neighbor and next-nearest-neighbor symmetric exchange coefficients. The signs and amplitudes of $J_1$ and $J_2$ provide important information about the magnetic phase of the system. A negative sign of $J_{1(2)}$ indicates FM coupling of the nearest-neighbor (next-nearest-neighbor) Cr atoms, while a positive sign indicates AFM coupling.

Figure \ref{J1graph}(a) shows that $J_1$ increases with increasing compressive strain up to a critical value of $7\%$, where the sign of $J_1$ changes from negative to positive. This phase transition from an FM phase to an AFM phase is consistent with the aforementioned positive $\Delta E$, as illustrated in Fig. \ref{bandgapforstrain} \cite{LucaWebprb}. Within the considered range of applied compressive strains, the next-nearest-neighbor exchange interaction $J_2$ remains negative, and its amplitude increases. This indicates that no magnetic frustration occurs in either the FM or AFM phase of the system.

Our results indicate that both $J_1$ and $J_2$ are almost unaffected by tensile strain, as shown in Fig. \ref{J1graph}(a).

\begin{figure}
\begin{center}
 \includegraphics*[width=8.4cm]{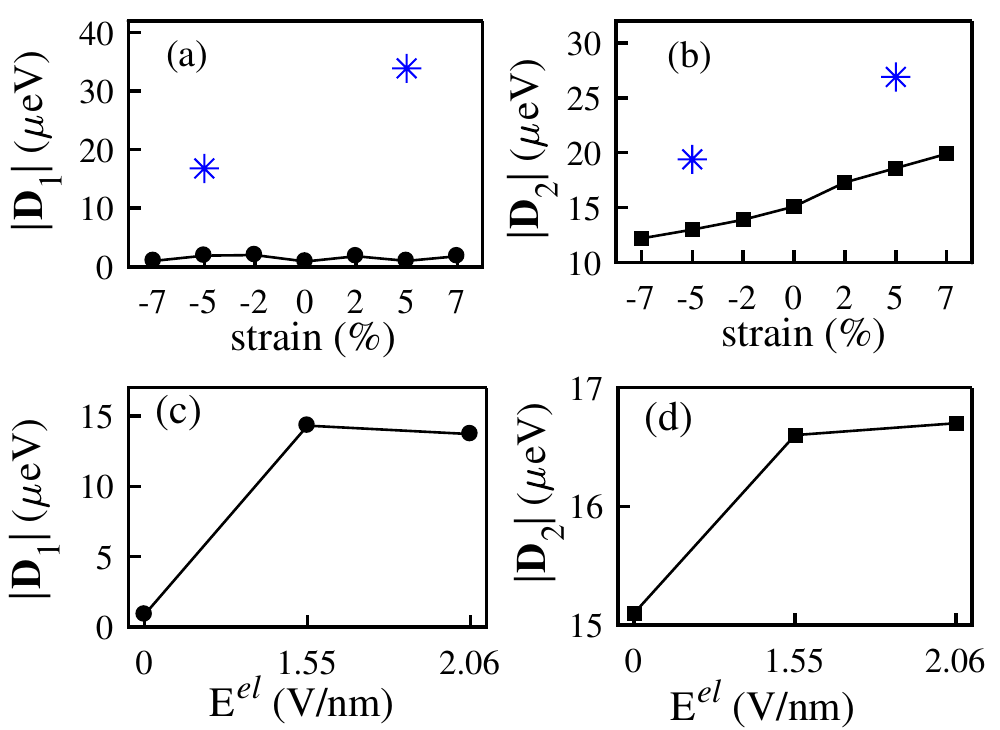}
\caption{ (Color online) (a) and (b) The nearest-neighbor and next-nearest-neighbor DM interactions of monolayer CrI$_3$ versus biaxial strain. The star symbols indicate the values of $|{\bf D_1}|$ and $|{\bf D_2}|$ under $5 \%$ compressive and tensile uniaxial strains. (c) and (d) The nearest-neighbor and next-nearest-neighbor DM interactions as functions of the perpendicular electric field.}\label{fig:8}
 \label{DMIstrain}
\end{center}
\end{figure}

\begin{table*}
\caption{The effects of biaxial strains on the nearest-neighbor and next-nearest-neighbor DM vectors of monolayer CrI$_3$.}
\begin{center}
\begin{tabular}{cccccc}
  \hline
  \hline
   strain & ${\bf D}_1$ ($\mu$eV) & $|{\bf D}_1|$ ($\mu$eV) & ${\bf D}_2 $~($\mu$eV)  &  $|{\bf D}_2|$ ($\mu$eV)  \\
  \hline
  $-7 \%$ &  (0.9,-0.2,-0.4) & 1.0 & (7.1, 5.8, 8.1) &  12.2  \\
  \hline
  $-5 \%$ &  (-0.5, 1.8, -0.1 ) & 1.9  & (8.2, 4.2, 9.2)  & 13.0   \\
  \hline
  $-2 \%$ &  (0.00, 1.8, 0.9) & 2.0 &  (8.4, 4.9, 9.9) & 13.9  \\
  \hline
  $ 0 \%$ &  (-0.2, 0.8, 0.3) & 0.9 & (9.6, -2.7, 11.3) & 15.1    \\
  \hline
  $+2 \%$ &   (-0.8, 1.5, 0.5) & 1.8 & (10.0, 5.4, 13.1)  & 17.3  \\
  \hline
  $+5 \%$ &   (0.00, 1.0, 0.3 ) &1.0 &   (10.1, 6.0, 14.4)   & 18.6   \\
  \hline
  $+7 \%$ &  (0.3, 1.7, 0.4)& 1.8 & (10.3, -6.3, -15.8)  & 19.9  \\
  \hline
  \hline
\end{tabular}
\end{center}
\label{tabelofJ_1andMAEforstrain}
\end{table*}

\begin{table*}[!t]
\caption{The effects of uniaxial strains on the nearest-neighbor and next-nearest-neighbor DM vectors in monolayer CrI$_3$. }
\begin{center}
\begin{tabular}{cccccc}
  \hline
  \hline
   strain & ${\bf D}_1$ ($\mu$eV) & $|{\bf D}_1|$ ($\mu$eV) & ${\bf D}_2$ ($\mu$eV)  &  $|{\bf D}_2|$ ($\mu$eV)  \\
  \hline
  $-5 \%$ &  (-10.2, -8.2, -10.7) & 16.8  & (15.9, 11.0, 1.7)  & 19.4   \\
  \hline
  $ 0 \%$ &  (-0.2, 0.8, 0.3) & 0.9 & (0.9.6, -2.7, 11.3) & 15.1    \\
  \hline
  $+5 \%$ &   (-0.3, 10.5, 32.2) &33.9 &   (11.4, 0.8, 24.4)  & 26.9   \\
  \hline
  \hline
\end{tabular}
\end{center}
\label{tabelDMIuni_st}
\end{table*}

\begin{table*}
\caption{The effects of perpendicular electric fields on the nearest-neighbor and next-nearest-neighbor DM vectors in monolayer CrI$_3$. }
\begin{center}
\begin{tabular}{ccccc}
  \hline
  \hline
        electric field (V/nm) & ${\bf D}_1$ ($\mu$eV) & $|{\bf D}_1|$ ($\mu$eV) & ${\bf D}_2$ ($\mu$eV) & $ |{\bf D}_2|$ ($\mu$eV) \\

  \hline
   0 & (-0.2, 0.8, 0.3) & 0.9 & (9.6, -2.7, 11.3) &   15.1   \\
  \hline
   1.55 (V/nm) & (-0.2, 1.6, -14.2) & 14.3 & (-9.4, -5.4, 12.6)  & 16.6  \\
  \hline
   2.06 (V/nm) & (-1.3, 1.0, -13.6) & 13.7 &(10.2, 6.6, 11.4)  & 16.7    \\
  \hline
  \hline
\end{tabular}
\end{center}
\label{TabelofDMI}
\end{table*}

The MAE of CrI$_3$ is found to be approximately $-0.54$ meV per Cr for a pristine monolayer, as illustrated in Fig. \ref{J1graph}(b)), in agreement with previous reports \cite{Liu2018,LucaWebprb}. This tells us that monolayer CrI$_3$ is a material with perpendicular magnetic anisotropy.

Figure \ref{J1graph}(b) shows that the sign of the MAE does not change under an applied strain; thus, the magnetic direction of the system continues to point out of the plane. However, the strength of the MAE is significantly changed by a compressive strain. A compressive strain increases the MAE by more than $300\%$. In a 2D magnetic material, the MAE determines the critical magnetic temperature, being the Curie or N\'eel temperature in an FM or AFM material, respectively. Consequently, the phase transition temperature of monolayer CrI$_3$ can be dramatically increased by applying a strain.

Finally, we compute the nearest-neighbor and next-nearest-neighbor DM vectors in monolayer CrI$_3$. We find that the amplitude of ${\bf{D}}_1$ is at least one order of magnitude smaller than that of ${\bf{D}}_2$ in both the absence and presence of strain.
For a pristine layer, we find that $|{{\bf D}_1}|= 0.9$ $\mu$eV and $|{{\bf D}_2}|= 15.1$ $\mu$eV. Our results also show that the direction of ${\bf{D}}_2$ is not exactly perpendicular to the plane, as theoretically predicted for an ideal hexagonal lattice \cite{Alireza}; instead, it deviates from the $z$-direction. Figure \ref{DMIstrain} and Table \ref{tabelofJ_1andMAEforstrain} demonstrate that strain can be applied to control both the amplitudes and directions of the DM vectors.
Figure \ref{DMIstrain} shows that $|{{\bf D}_1}|$ does not markedly change, while a biaxial tensile (compressive) strain causes $|{{\bf D}_2}|$ to increase (decrease) by more than $30 \%$.

Since the values of the DM interactions are one order of magnitude smaller than the magnetic anisotropy throughout the entire range of applied strains considered here, the spins in the system remain collinear, and no chiral ground state emerges in this monolayer \cite{Hxxz,Liu2018,Ghosh}.
However, the DM interactions modify the magnon dispersion; thus, they can be experimentally measured by using techniques such as
the magneto-optical Kerr probe technique \cite{Kerr}, magneto-Raman spectroscopy \cite{Raman},
Brillouin light scattering \cite{nembach2015linear,di2015direct,belmeguenai2015interfacial} or inelastic neutron scattering \cite{LebingChen, chen2020magnetic, INS-AFM, mena2014spin}.
Although our calculations show that the ground state of the system is not a chiral state, single chiral skyrmions as {\it{metastable}} states \cite{skyrmion1,skyrmion2} and chiral domain walls \cite{thiaville2012dynamics, qaiumzadeh2018controlling, ryu2013chiral} might be stably formed in this monolayer with proper tuning of the ratio between the DM interactions and the MAE.

\subsection{Effects of uniaxial strains on the electronic and magnetic properties}

With the application of a uniaxial strain to monolayer CrI$_3$, its geometric symmetries are altered, and as a consequence, the spin-spin interactions are modified. Here, we investigate the effects of compressive and tensile uniaxial strains on the magnetic properties of the monolayer. For this purpose, the unit cell of CrI$_3$ is either stretched or compressed along the $x$-direction, i.e., in the direction of the ${\bm{a}}$ lattice vector of the unit cell, as shown in Fig. \ref{geometryCrI3}(b).

Our calculations show that the signs of $J_1$ and $J_2$ remain negative under $5\%$ tensile and compressive uniaxial strains; therefore, the ground state remains as the FM state. The nearest-neighbor exchange coupling is reduced by approximately $3\%$ under a tensile uniaxial strain of $5\%$, while the corresponding compressive uniaxial strain causes a reduction of $25\%$. On the other hand, the next-nearest-neighbor exchange interaction is increased by approximately $12\%$ ($22\%$) by a uniaxial tensile (compressive) strain. The variations in $J_1$ and $J_2$ under uniaxial strains are similar to those under biaxial strains, as discussed in the previous section.

Under a uniaxial strain of $5 \%$, the MAE remains negative, and thus, the uniaxial anisotropy is in the out-of-plane direction. The MAE is increased by approximately $12 \%$ under a tensile uniaxial strain, while it becomes two times greater (-1.06 meV per Cr) under a compressive uniaxial strain.

Table \ref{tabelDMIuni_st} presents the nearest-neighbor and next-nearest-neighbor DM vectors of monolayer CrI$_3$ under different uniaxial strains. A uniaxial strain enhances the nearest-neighbor DM interaction by breaking the inversion symmetry. Figure \ref{DMIstrain}(a) shows that a uniaxial strain causes $|{\bf D_1}|$ to increase by an order of magnitude compared to its value under a biaxial strain.
Moreover, a uniaxial strain also enhances $|{\bf D_2}|$ more efficiently than a biaxial strain does (see Fig. \ref{DMIstrain}(b)).
In addition, the directions of the DM vectors are strongly affected by the application of a uniaxial strain, as shown in Table \ref{tabelDMIuni_st}.

\subsection{Effects of perpendicular electric fields}

The effects of electric fields on the magnetic properties of monolayer CrI$_3$ are fascinating, and some properties of CrI$_3$ under an external electric field have already been explored recently \cite{Behera,Liu2018,Ghosh}. Here, we apply a saw-like potential with dipole correction to establish a uniform electric field in the $z$-direction across the monolayer CrI$_3$.

The FM state remains the more stable configuration in the presence of an electric field. The bandgap of the monolayer CrI$_3$ slowly decreases with an increasing electric field. Our results, shown in Fig. \ref{J1graph}(c), indicate that an electric field leads to an increase in both the nearest-neighbor and next-nearest-neighbor exchange interactions, but the sign remains negative, in agreement with previous studies \cite{Behera,Liu2018}. The out-of-plane MAE also increases (becomes more negative) under an electric field, as shown in Fig. \ref{J1graph}(d).

The effects of perpendicular electric fields on the amplitudes and directions of the DM vectors are presented in Figs. \ref{DMIstrain}(c) and (d) and Table \ref{TabelofDMI}. A perpendicular electric field breaks the inversion symmetry and thus causes the nearest-neighbor DM interaction to increase dramatically, from almost zero to more than $10$~$\mu eV$. Additionally, since the SOC is increased under the application of a perpendicular electric field, the next-nearest-neighbor DM interaction is also increased by approximately $10\%$.


\section{Summary and concluding remarks}
The spin-spin interactions determine the magnetic phases and critical phase transition temperatures of 2D magnetic systems.
Therefore, methods for the efficient control and manipulation of these interactions are essential for utilizing these materials in novel functional spintronic devices.

In this work, monolayer CrI$_3$, as a representative 2D magnetic material, has been investigated using spin-dependent DFT. We have studied the electronic and magnetic ground states of this monolayer under biaxial and uniaxial mechanical strains as well as a perpendicular electric field. By mapping the first-principle DFT solutions onto a spin Hamiltonian, we have derived the relevant spin-spin interaction parameters. 

We have shown that the sign and amplitude of the nearest-neighbor exchange interaction can be significantly modified and, thus, that a phase transition from the FM phase to the AFM phase is possible under a suitable compressive strain. On the other hand, a tensile strain or an electric field can affect only the amplitude of the nearest-neighbor exchange interaction. In all cases, the sign of the next-nearest-neighbor exchange interaction remains negative, indicating that the system is a collinear magnet and thus magnetically unfrustrated.

While the system remains uniaxial in the presence of both a strain and an electric field, the amplitude of the MAE changes dramatically. Consequently, the critical phase transition temperature can be enhanced, which is promising for room-temperature applications.

We have also examined the effects of strains and electric fields on the nearest-neighbor and next-nearest-neighbor DM interactions. We have shown that either a strain or an electric field can be used to control both the directions and amplitudes of the DM vectors. Thus, it is possible to design chiral spin textures and chiral magnon transport in this 2D magnetic system by tuning the DM interactions.

Since several ab initio works on the magnetic properties of monolayer CrI$_3$ have been reported, a proper comparison with their results seems to be in order.
Zhang et al. \cite{ref14} considered a Heisenberg spin Hamiltonian and obtained the strain dependence of the first, second and third nearest neighbors exchange interaction parameters. Leon et al. \cite{leon2020strain} calculated only the in-plane Cr-Cr exchange coupling as a function of strain for CrI$_3$ monolayer and bilayer systems. Webster and Yan \cite{LucaWebprb} fitted DFT results with a Heisenberg spin model to evaluate the exchange interactions of the nearest neighbors as functions of the biaxial strain. An XXZ model Hamiltonian without DM interaction terms was also considered by Liu et al. \cite{liu2018multi} to compute the magnetic properties of a CrI$_3$ monolayer. Furthermore, $|{\bf D}_1|$ was reported only in the presence of an external electric field in \cite{Ghosh, Liu2018, Behera}. To summarize, we have extracted the pertinent parameters from the figures presented in those studies and compared their results with our own numerical results (see Tables VII and VIII).

The conclusion of these detailed comparisons is that
where the comparisons are appropriate, our results incorporate the complete magnetic parameters of the XXZ model with DM interaction. Thus, the present work yield more consistent contributions of the different magnetic parameters than previous works do.
Moreover, our results extend to several pertinent cases that have not been discussed before. Our findings here provide a broad outlook on future studies and potential applications of emerging magnetic 2D crystalline materials and, furthermore, can be explored using current experimental techniques.

\begin{table*}
\caption{A breakdown of the results for the values of the magnetic parameters as functions of the biaxial strain in a monolayer CrI$_3$ system. In some cases, the physical parameters were considered at different strains; therefore, we indicate the applied strains in superscript notation.}
\begin{center}
\begin{tabular}{ccccccccccccc}
\hline
\hline
strain &                              & -7 $\%$ & -5 $\%$& -2 $\%$ & 0       & +2 $\%$  & +5 $\%$  & +7 $\%$    \\
 \hline
       &         Present work                     & +0.76 & -1.34 & -2.98 & -3.91 & -4.15 & -3.96 & -3.91    \\
 $J_1$ & Ref. \cite{leon2020strain}   & ~~~~~~+2.34 $^{-6 \%}$  & +1.04 &  -1.83  & -2.85   & -3.18  &   -3.14  & -     \\
 (meV) & Ref. \cite{liu2018multi}    & -       & +0.11 &   -1.49 &  -1.96  & -1.99   &   -1.69  & -      \\
       & Ref. \cite{ref14}  & ~~~~~~+0.52 $^{-10 \%}$&~~~~~~-2.03 $^{-4 \%}$ &   -2.522& -2.94   &  -3.23   &~~~~~~-3.61 $^{+4 \%}$     &~~~~~~-3.94 $^{+6 \%}$   \\
       & Ref. \cite{LucaWebprb}     & ~~~~~~~+0.44 $^{-6.5 \%}$&~~~~~~~-1.11 $^{-4.5 \%}$ &~~~~~~~-2.07$ ^{-2.5 \%}$& -2.7 & ~~~~~~-2.37 $^{+3 \%}$ &~~~~~~~ -2.15$ ^{+4.5 \%}$ & ~~~~~~-1.93 $^{+6.5 \%}$\\
 \hline
 $J_2$ &     Present work                         & -0.93             &   -0.91            & -0.89  &  -0.67  & -0.69    &   -0.72               &  -0.70           \\
 (meV) & Ref. \cite{ref14}  & ~~~~~~-0.15 $^{-10 \%}$&~~~~~~ -0.68 $^{-4 \%}$ & -0.65  & -0.64   &  -0.58   & ~~~~~~-0.54 $^{+4 \%}$     &~~~~~~ -0.54 $^{+6 \%}$   \\
 \hline
       &        Present work                      &  -1.89            &    -1.38           &  -1.05 & -0.54   &   -0.17  &  -0.84                &  -0.17           \\
  MAE  & Ref. \cite{LucaWebprb}       &~~~~~~-1.89 $^{-8 \%}$   &~~~~~~~-1.10 $^{-4.5 \%}$  & ~~~~~~-0.87$ ^{-1 \%}$& -0.80 &~~~~~~~-0.70 $^{+2.5 \%}$ & ~~~~~~~-0.69$ ^{+6.5 \%}$ & ~~~~~~-0.63 $^{+10 \%}$\\
 (meV) & Ref. \cite{liu2018multi}    & -       & -0.74 &   -0.88 &  -0.82     &   -0.68  &   -0.41  & -      \\
       & Ref. \cite{leon2020strain}   & ~~~~~~-1.5 $^{-6 \%}$  & - &  -  & -0.68   & -  &   - & ~~~~~~-0.32 $^{+6 \%}$     \\
 \hline
 \hline
\end{tabular}
\end{center}
\label{compJstrain}
\end{table*}

\begin{table*}
\caption{A breakdown of the results for the values of the magnetic parameters under different electric fields in a monolayer CrI$_3$ system. In our work (other references), $E^{el}_1$=1.55 (1) V/nm and $E^{el}_1$=2.06 (2) V/nm.}
\begin{center}
\begin{tabular}{ccccccccccccc}
\hline
\hline
electric field  & &$J_1$ (meV)         & ~~~~~~~$J_2$ (meV)~~~~~~~~~~&  & MAE (meV)          &                &     &   &$|\bf{D_1}|$  ($\mu$eV)         &                  &  \\
                      &    & Ref.~\cite{Liu2018}&                          &   & Ref.~\cite{Liu2018} &  Ref. \cite{Behera}& ~~~~~~~~~~ &  &Ref.~\cite{Ghosh}& Ref.~\cite{Liu2018} & Ref.~\cite{Behera}\\
 \hline
0                     & -3.91&  -1.96           & -0.49       & -0.54&  -0.80   &  -   &  &0.9  &  0  &  0   &  0 \\
 \hline
$E^{el}_1$            & -4.46&  -2.0            & -0.74       & -0.55&  -0.75   &  -   &  &14.3 & 20  & 400  &  -  \\
 \hline
$E^{el}_2$            &-4.58 &  -2.1         & -0.78       & -0.55&  -0.475  &  0.51 &  &13.7 &  4  & 800  & 180  \\
 \hline
 \hline
\end{tabular}
\end{center}
\label{compJEl}
\end{table*}

\begin{acknowledgments}
We thank Ali Ebrahimian for fruitful discussions. This work was supported by the Iran Science Elites Federation.
A. Q. was supported by the European Research Council via Advanced Grant No. 669442, ``Insulatronics,'' and by the Research Council of Norway through its Centres of Excellence funding scheme, Project No. 262633, ``QuSpin.''
\end{acknowledgments}

\nocite{apsrev41Control}
\bibliographystyle{apsrev4-1}
\bibliography{saha}

\begin{thebibliography}{60}%
\makeatletter
\providecommand \@ifxundefined [1]{%
 \@ifx{#1\undefined}
}%
\providecommand \@ifnum [1]{%
 \ifnum #1\expandafter \@firstoftwo
 \else \expandafter \@secondoftwo
 \fi
}%
\providecommand \@ifx [1]{%
 \ifx #1\expandafter \@firstoftwo
 \else \expandafter \@secondoftwo
 \fi
}%
\providecommand \natexlab [1]{#1}%
\providecommand \enquote  [1]{``#1''}%
\providecommand \bibnamefont  [1]{#1}%
\providecommand \bibfnamefont [1]{#1}%
\providecommand \citenamefont [1]{#1}%
\providecommand \href@noop [0]{\@secondoftwo}%
\providecommand \href [0]{\begingroup \@sanitize@url \@href}%
\providecommand \@href[1]{\@@startlink{#1}\@@href}%
\providecommand \@@href[1]{\endgroup#1\@@endlink}%
\providecommand \@sanitize@url [0]{\catcode `\\12\catcode `\$12\catcode
  `\&12\catcode `\#12\catcode `\^12\catcode `\_12\catcode `\%12\relax}%
\providecommand \@@startlink[1]{}%
\providecommand \@@endlink[0]{}%
\providecommand \url  [0]{\begingroup\@sanitize@url \@url }%
\providecommand \@url [1]{\endgroup\@href {#1}{\urlprefix }}%
\providecommand \urlprefix  [0]{URL }%
\providecommand \Eprint [0]{\href }%
\providecommand \doibase [0]{http://dx.doi.org/}%
\providecommand \selectlanguage [0]{\@gobble}%
\providecommand \bibinfo  [0]{\@secondoftwo}%
\providecommand \bibfield  [0]{\@secondoftwo}%
\providecommand \translation [1]{[#1]}%
\providecommand \BibitemOpen [0]{}%
\providecommand \bibitemStop [0]{}%
\providecommand \bibitemNoStop [0]{.\EOS\space}%
\providecommand \EOS [0]{\spacefactor3000\relax}%
\providecommand \BibitemShut  [1]{\csname bibitem#1\endcsname}%
\let\auto@bib@innerbib\@empty
\bibitem [{\citenamefont {Novoselov}\ \emph {et~al.}(2004)\citenamefont
  {Novoselov}, \citenamefont {Geim}, \citenamefont {Morozov}, \citenamefont
  {Jiang}, \citenamefont {Zhang}, \citenamefont {Dubonos}, \citenamefont
  {Grigorieva},\ and\ \citenamefont {Firsov}}]{ref1}%
  \BibitemOpen
  \bibfield  {author} {\bibinfo {author} {\bibfnamefont {K.~S.}\ \bibnamefont
  {Novoselov}}, \bibinfo {author} {\bibfnamefont {A.~K.}\ \bibnamefont {Geim}},
  \bibinfo {author} {\bibfnamefont {S.~V.}\ \bibnamefont {Morozov}}, \bibinfo
  {author} {\bibfnamefont {D.}~\bibnamefont {Jiang}}, \bibinfo {author}
  {\bibfnamefont {Y.}~\bibnamefont {Zhang}}, \bibinfo {author} {\bibfnamefont
  {S.~V.}\ \bibnamefont {Dubonos}}, \bibinfo {author} {\bibfnamefont {I.~V.}\
  \bibnamefont {Grigorieva}}, \ and\ \bibinfo {author} {\bibfnamefont {A.~A.}\
  \bibnamefont {Firsov}},\ }\bibfield  {title} {\enquote {\bibinfo {title}
  {{Electric field effect in atomically thin carbon films}},}\ }\href {\doibase
  10.1126/science.1102896} {\bibfield  {journal} {\bibinfo  {journal}
  {Science}\ }\textbf {\bibinfo {volume} {306}},\ \bibinfo {pages} {666}
  (\bibinfo {year} {2004})}\BibitemShut {NoStop}%
\bibitem [{\citenamefont {Huang}\ \emph {et~al.}(2017)\citenamefont {Huang},
  \citenamefont {Clark}, \citenamefont {Navarro-Moratalla}, \citenamefont
  {Klein}, \citenamefont {Cheng}, \citenamefont {Seyler}, \citenamefont
  {Zhong}, \citenamefont {Schmidgall}, \citenamefont {McGuire}, \citenamefont
  {Cobden} \emph {et~al.}}]{huang2017layer}%
  \BibitemOpen
  \bibfield  {author} {\bibinfo {author} {\bibfnamefont {B.}~\bibnamefont
  {Huang}}, \bibinfo {author} {\bibfnamefont {G.}~\bibnamefont {Clark}},
  \bibinfo {author} {\bibfnamefont {E.}~\bibnamefont {Navarro-Moratalla}},
  \bibinfo {author} {\bibfnamefont {D.~R.}\ \bibnamefont {Klein}}, \bibinfo
  {author} {\bibfnamefont {R.}~\bibnamefont {Cheng}}, \bibinfo {author}
  {\bibfnamefont {K.~L.}\ \bibnamefont {Seyler}}, \bibinfo {author}
  {\bibfnamefont {D.}~\bibnamefont {Zhong}}, \bibinfo {author} {\bibfnamefont
  {E.}~\bibnamefont {Schmidgall}}, \bibinfo {author} {\bibfnamefont {M.~A.}\
  \bibnamefont {McGuire}}, \bibinfo {author} {\bibfnamefont {D.~H.}\
  \bibnamefont {Cobden}},  \emph {et~al.},\ }\bibfield  {title} {\enquote
  {\bibinfo {title} {{Layer-dependent ferromagnetism in a van der Waals crystal
  down to the monolayer limit}},}\ }\href {\doibase 10.1038/nature22391}
  {\bibfield  {journal} {\bibinfo  {journal} {Nature}\ }\textbf {\bibinfo
  {volume} {546}},\ \bibinfo {pages} {270} (\bibinfo {year}
  {2017})}\BibitemShut {NoStop}%
\bibitem [{\citenamefont {Mermin}\ and\ \citenamefont {Wagner}(1966)}]{ref2}%
  \BibitemOpen
  \bibfield  {author} {\bibinfo {author} {\bibfnamefont {N.~D.}\ \bibnamefont
  {Mermin}}\ and\ \bibinfo {author} {\bibfnamefont {H.}~\bibnamefont
  {Wagner}},\ }\bibfield  {title} {\enquote {\bibinfo {title} {{Absence of
  ferromagnetism or antiferromagnetism in one- or two-dimensional isotropic
  Heisenberg models}},}\ }\href {\doibase 10.1103/PhysRevLett.17.1133}
  {\bibfield  {journal} {\bibinfo  {journal} {Phys. Rev. Lett.}\ }\textbf
  {\bibinfo {volume} {17}},\ \bibinfo {pages} {1133} (\bibinfo {year}
  {1966})}\BibitemShut {NoStop}%
\bibitem [{\citenamefont {Feng}\ \emph {et~al.}(2017)\citenamefont {Feng},
  \citenamefont {Shen}, \citenamefont {Yang}, \citenamefont {Wang},
  \citenamefont {Zeng}, \citenamefont {Wu}, \citenamefont {Chintalapati},\ and\
  \citenamefont {Chang}}]{2Dreview1}%
  \BibitemOpen
  \bibfield  {author} {\bibinfo {author} {\bibfnamefont {Y.~P.}\ \bibnamefont
  {Feng}}, \bibinfo {author} {\bibfnamefont {L.}~\bibnamefont {Shen}}, \bibinfo
  {author} {\bibfnamefont {M.}~\bibnamefont {Yang}}, \bibinfo {author}
  {\bibfnamefont {A.}~\bibnamefont {Wang}}, \bibinfo {author} {\bibfnamefont
  {M.}~\bibnamefont {Zeng}}, \bibinfo {author} {\bibfnamefont {Q.}~\bibnamefont
  {Wu}}, \bibinfo {author} {\bibfnamefont {S.}~\bibnamefont {Chintalapati}}, \
  and\ \bibinfo {author} {\bibfnamefont {C.}~\bibnamefont {Chang}},\ }\bibfield
   {title} {\enquote {\bibinfo {title} {{Prospects of spintronics based on 2D
  materials}},}\ }\href {\doibase 10.1002/wcms.1313} {\bibfield  {journal}
  {\bibinfo  {journal} {Wiley Interdiscip. Rev. Comput. Mol. Sci.}\ }\textbf
  {\bibinfo {volume} {7}},\ \bibinfo {pages} {e1313} (\bibinfo {year}
  {2017})}\BibitemShut {NoStop}%
\bibitem [{\citenamefont {Mounet}\ \emph {et~al.}(2018)\citenamefont {Mounet},
  \citenamefont {Gibertini}, \citenamefont {Schwaller}, \citenamefont {Campi},
  \citenamefont {Merkys}, \citenamefont {Marrazzo}, \citenamefont {Sohier},
  \citenamefont {Castelli}, \citenamefont {Eligio}, \citenamefont {Cepellotti},
  \citenamefont {Pizzi},\ and\ \citenamefont {Marzari}}]{2Dreview2}%
  \BibitemOpen
  \bibfield  {author} {\bibinfo {author} {\bibfnamefont {N.}~\bibnamefont
  {Mounet}}, \bibinfo {author} {\bibfnamefont {M.}~\bibnamefont {Gibertini}},
  \bibinfo {author} {\bibfnamefont {P.}~\bibnamefont {Schwaller}}, \bibinfo
  {author} {\bibfnamefont {D.}~\bibnamefont {Campi}}, \bibinfo {author}
  {\bibfnamefont {A.}~\bibnamefont {Merkys}}, \bibinfo {author} {\bibfnamefont
  {A.}~\bibnamefont {Marrazzo}}, \bibinfo {author} {\bibfnamefont
  {T.}~\bibnamefont {Sohier}}, \bibinfo {author} {\bibnamefont {Castelli}},
  \bibinfo {author} {\bibfnamefont {I.}~\bibnamefont {Eligio}}, \bibinfo
  {author} {\bibfnamefont {A.}~\bibnamefont {Cepellotti}}, \bibinfo {author}
  {\bibfnamefont {G.}~\bibnamefont {Pizzi}}, \ and\ \bibinfo {author}
  {\bibfnamefont {N.}~\bibnamefont {Marzari}},\ }\bibfield  {title} {\enquote
  {\bibinfo {title} {{Two-dimensional materials from high-throughput
  computational exfoliation of experimentally known compounds}},}\ }\href
  {\doibase 10.1038/s41565-017-0035-5} {\bibfield  {journal} {\bibinfo
  {journal} {Nat. Nanotechnol.}\ }\textbf {\bibinfo {volume} {13}},\ \bibinfo
  {pages} {246} (\bibinfo {year} {2018})}\BibitemShut {NoStop}%
\bibitem [{\citenamefont {Gibertini}\ \emph {et~al.}(2019)\citenamefont
  {Gibertini}, \citenamefont {Koperski}, \citenamefont {Morpurgo},\ and\
  \citenamefont {Novoselov}}]{2Dreview3}%
  \BibitemOpen
  \bibfield  {author} {\bibinfo {author} {\bibfnamefont {M.}~\bibnamefont
  {Gibertini}}, \bibinfo {author} {\bibfnamefont {M.}~\bibnamefont {Koperski}},
  \bibinfo {author} {\bibfnamefont {A.~F.}\ \bibnamefont {Morpurgo}}, \ and\
  \bibinfo {author} {\bibfnamefont {K.~S.}\ \bibnamefont {Novoselov}},\
  }\bibfield  {title} {\enquote {\bibinfo {title} {{Magnetic 2D materials and
  heterostructures}},}\ }\href {\doibase 10.1038/s41565-019-0438-6} {\bibfield
  {journal} {\bibinfo  {journal} {Nat. Nanotechnol.}\ }\textbf {\bibinfo
  {volume} {14}},\ \bibinfo {pages} {408} (\bibinfo {year} {2019})}\BibitemShut
  {NoStop}%
\bibitem [{\citenamefont {Wang}\ \emph {et~al.}(2020)\citenamefont {Wang},
  \citenamefont {Huang}, \citenamefont {Cheung}, \citenamefont {Chen},
  \citenamefont {Tan}, \citenamefont {Huang}, \citenamefont {Zhao},
  \citenamefont {Zhao}, \citenamefont {Wu}, \citenamefont {Feng}, \citenamefont
  {Wu},\ and\ \citenamefont {Chang}}]{2Dreview4}%
  \BibitemOpen
  \bibfield  {author} {\bibinfo {author} {\bibfnamefont {M.-C.}\ \bibnamefont
  {Wang}}, \bibinfo {author} {\bibfnamefont {C.-C.}\ \bibnamefont {Huang}},
  \bibinfo {author} {\bibfnamefont {C.-H.}\ \bibnamefont {Cheung}}, \bibinfo
  {author} {\bibfnamefont {C.-Y.}\ \bibnamefont {Chen}}, \bibinfo {author}
  {\bibfnamefont {S.~G.}\ \bibnamefont {Tan}}, \bibinfo {author} {\bibfnamefont
  {T.-W.}\ \bibnamefont {Huang}}, \bibinfo {author} {\bibfnamefont
  {Y.}~\bibnamefont {Zhao}}, \bibinfo {author} {\bibfnamefont {Y.}~\bibnamefont
  {Zhao}}, \bibinfo {author} {\bibfnamefont {G.}~\bibnamefont {Wu}}, \bibinfo
  {author} {\bibfnamefont {Y.-P.}\ \bibnamefont {Feng}}, \bibinfo {author}
  {\bibfnamefont {H.-C.}\ \bibnamefont {Wu}}, \ and\ \bibinfo {author}
  {\bibfnamefont {C.-R.}\ \bibnamefont {Chang}},\ }\bibfield  {title} {\enquote
  {\bibinfo {title} {{Prospects and opportunities of 2D van der Waals magnetic
  systems}},}\ }\href {\doibase 10.1002/andp.201900452} {\bibfield  {journal}
  {\bibinfo  {journal} {Ann. Phys. (Berlin)}\ ,\ \bibinfo {pages} {1900452}}
  (\bibinfo {year} {2020})}\BibitemShut {NoStop}%
\bibitem [{\citenamefont {Burch}\ \emph {et~al.}(2018)\citenamefont {Burch},
  \citenamefont {Mandrus},\ and\ \citenamefont {Park}}]{burch2018magnetism}%
  \BibitemOpen
  \bibfield  {author} {\bibinfo {author} {\bibfnamefont {K.~S.}\ \bibnamefont
  {Burch}}, \bibinfo {author} {\bibfnamefont {D.}~\bibnamefont {Mandrus}}, \
  and\ \bibinfo {author} {\bibfnamefont {J.-G.}\ \bibnamefont {Park}},\
  }\bibfield  {title} {\enquote {\bibinfo {title} {{Magnetism in
  two-dimensional van der Waals materials}},}\ }\href {\doibase
  10.1038/s41586-018-0631-z} {\bibfield  {journal} {\bibinfo  {journal}
  {Nature}\ }\textbf {\bibinfo {volume} {563}},\ \bibinfo {pages} {47}
  (\bibinfo {year} {2018})}\BibitemShut {NoStop}%
\bibitem [{\citenamefont {Gong}\ and\ \citenamefont
  {Zhang}(2019)}]{gong2019two}%
  \BibitemOpen
  \bibfield  {author} {\bibinfo {author} {\bibfnamefont {C.}~\bibnamefont
  {Gong}}\ and\ \bibinfo {author} {\bibfnamefont {X.}~\bibnamefont {Zhang}},\
  }\bibfield  {title} {\enquote {\bibinfo {title} {{Two-dimensional magnetic
  crystals and emergent heterostructure devices}},}\ }\href {\doibase
  10.1126/science.aav4450} {\bibfield  {journal} {\bibinfo  {journal}
  {Science}\ }\textbf {\bibinfo {volume} {363}} (\bibinfo {year} {2019}),\
  10.1126/science.aav4450}\BibitemShut {NoStop}%
\bibitem [{\citenamefont {Klein}\ \emph {et~al.}(2018)\citenamefont {Klein},
  \citenamefont {MacNeill}, \citenamefont {Lado}, \citenamefont {Soriano},
  \citenamefont {Navarro-Moratalla}, \citenamefont {Watanabe}, \citenamefont
  {Taniguchi}, \citenamefont {Manni}, \citenamefont {Canfield}, \citenamefont
  {Fern{\'a}ndez-Rossier} \emph {et~al.}}]{klein2018probing}%
  \BibitemOpen
  \bibfield  {author} {\bibinfo {author} {\bibfnamefont {D.~R.}\ \bibnamefont
  {Klein}}, \bibinfo {author} {\bibfnamefont {D.}~\bibnamefont {MacNeill}},
  \bibinfo {author} {\bibfnamefont {J.~L.}\ \bibnamefont {Lado}}, \bibinfo
  {author} {\bibfnamefont {D.}~\bibnamefont {Soriano}}, \bibinfo {author}
  {\bibfnamefont {E.}~\bibnamefont {Navarro-Moratalla}}, \bibinfo {author}
  {\bibfnamefont {K.}~\bibnamefont {Watanabe}}, \bibinfo {author}
  {\bibfnamefont {T.}~\bibnamefont {Taniguchi}}, \bibinfo {author}
  {\bibfnamefont {S.}~\bibnamefont {Manni}}, \bibinfo {author} {\bibfnamefont
  {P.}~\bibnamefont {Canfield}}, \bibinfo {author} {\bibfnamefont
  {J.}~\bibnamefont {Fern{\'a}ndez-Rossier}},  \emph {et~al.},\ }\bibfield
  {title} {\enquote {\bibinfo {title} {{Probing magnetism in 2D van der Waals
  crystalline insulators via electron tunneling}},}\ }\href {\doibase
  10.1126/science.aar3617} {\bibfield  {journal} {\bibinfo  {journal}
  {Science}\ }\textbf {\bibinfo {volume} {360}},\ \bibinfo {pages} {1218}
  (\bibinfo {year} {2018})}\BibitemShut {NoStop}%
\bibitem [{\citenamefont {Wang}\ \emph {et~al.}(2018)\citenamefont {Wang},
  \citenamefont {Guti{\'e}rrez-Lezama}, \citenamefont {Ubrig}, \citenamefont
  {Kroner}, \citenamefont {Gibertini}, \citenamefont {Taniguchi}, \citenamefont
  {Watanabe}, \citenamefont {Imamo{\u{g}}lu}, \citenamefont {Giannini},\ and\
  \citenamefont {Morpurgo}}]{wang2018very}%
  \BibitemOpen
  \bibfield  {author} {\bibinfo {author} {\bibfnamefont {Z.}~\bibnamefont
  {Wang}}, \bibinfo {author} {\bibfnamefont {I.}~\bibnamefont
  {Guti{\'e}rrez-Lezama}}, \bibinfo {author} {\bibfnamefont {N.}~\bibnamefont
  {Ubrig}}, \bibinfo {author} {\bibfnamefont {M.}~\bibnamefont {Kroner}},
  \bibinfo {author} {\bibfnamefont {M.}~\bibnamefont {Gibertini}}, \bibinfo
  {author} {\bibfnamefont {T.}~\bibnamefont {Taniguchi}}, \bibinfo {author}
  {\bibfnamefont {K.}~\bibnamefont {Watanabe}}, \bibinfo {author}
  {\bibfnamefont {A.}~\bibnamefont {Imamo{\u{g}}lu}}, \bibinfo {author}
  {\bibfnamefont {E.}~\bibnamefont {Giannini}}, \ and\ \bibinfo {author}
  {\bibfnamefont {A.~F.}\ \bibnamefont {Morpurgo}},\ }\bibfield  {title}
  {\enquote {\bibinfo {title} {{Very large tunneling magnetoresistance in
  layered magnetic semiconductor ${\mathrm{CrI}}_{3}$}},}\ }\href {\doibase
  10.1038/s41467-018-04953-8} {\bibfield  {journal} {\bibinfo  {journal} {Nat.
  commun.}\ }\textbf {\bibinfo {volume} {9}},\ \bibinfo {pages} {1} (\bibinfo
  {year} {2018})}\BibitemShut {NoStop}%
\bibitem [{\citenamefont {Huang}\ \emph {et~al.}(2018)\citenamefont {Huang},
  \citenamefont {Clark}, \citenamefont {Klein}, \citenamefont {MacNeill},
  \citenamefont {Navarro-Moratalla}, \citenamefont {Seyler}, \citenamefont
  {Wilson}, \citenamefont {McGuire}, \citenamefont {Cobden}, \citenamefont
  {Xiao} \emph {et~al.}}]{huang2018electrical}%
  \BibitemOpen
  \bibfield  {author} {\bibinfo {author} {\bibfnamefont {B.}~\bibnamefont
  {Huang}}, \bibinfo {author} {\bibfnamefont {G.}~\bibnamefont {Clark}},
  \bibinfo {author} {\bibfnamefont {D.~R.}\ \bibnamefont {Klein}}, \bibinfo
  {author} {\bibfnamefont {D.}~\bibnamefont {MacNeill}}, \bibinfo {author}
  {\bibfnamefont {E.}~\bibnamefont {Navarro-Moratalla}}, \bibinfo {author}
  {\bibfnamefont {K.~L.}\ \bibnamefont {Seyler}}, \bibinfo {author}
  {\bibfnamefont {N.}~\bibnamefont {Wilson}}, \bibinfo {author} {\bibfnamefont
  {M.~A.}\ \bibnamefont {McGuire}}, \bibinfo {author} {\bibfnamefont {D.~H.}\
  \bibnamefont {Cobden}}, \bibinfo {author} {\bibfnamefont {D.}~\bibnamefont
  {Xiao}},  \emph {et~al.},\ }\bibfield  {title} {\enquote {\bibinfo {title}
  {{Electrical control of 2D magnetism in bilayer ${\mathrm{CrI}}_{3}$}},}\
  }\href {\doibase 10.1038/s41565-018-0121-3} {\bibfield  {journal} {\bibinfo
  {journal} {Nat. nanotechnol.}\ }\textbf {\bibinfo {volume} {13}},\ \bibinfo
  {pages} {544} (\bibinfo {year} {2018})}\BibitemShut {NoStop}%
\bibitem [{\citenamefont {Wang}\ \emph {et~al.}(2016)\citenamefont {Wang},
  \citenamefont {Fan}, \citenamefont {Zhu},\ and\ \citenamefont
  {Wu}}]{wang2016doping}%
  \BibitemOpen
  \bibfield  {author} {\bibinfo {author} {\bibfnamefont {H.}~\bibnamefont
  {Wang}}, \bibinfo {author} {\bibfnamefont {F.}~\bibnamefont {Fan}}, \bibinfo
  {author} {\bibfnamefont {S.}~\bibnamefont {Zhu}}, \ and\ \bibinfo {author}
  {\bibfnamefont {H.}~\bibnamefont {Wu}},\ }\bibfield  {title} {\enquote
  {\bibinfo {title} {{Doping enhanced ferromagnetism and induced
  half-metallicity in ${\mathrm{CrI}}_{3}$ monolayer}},}\ }\href {\doibase
  10.1209/0295-5075/114/47001} {\bibfield  {journal} {\bibinfo  {journal} {EPL
  (Europhysics Letters)}\ }\textbf {\bibinfo {volume} {114}},\ \bibinfo {pages}
  {47001} (\bibinfo {year} {2016})}\BibitemShut {NoStop}%
\bibitem [{\citenamefont {Sivadas}\ \emph {et~al.}(2018)\citenamefont
  {Sivadas}, \citenamefont {Okamoto}, \citenamefont {Xu}, \citenamefont
  {Fennie},\ and\ \citenamefont {Xiao}}]{sivadas2018stacking}%
  \BibitemOpen
  \bibfield  {author} {\bibinfo {author} {\bibfnamefont {N.}~\bibnamefont
  {Sivadas}}, \bibinfo {author} {\bibfnamefont {S.}~\bibnamefont {Okamoto}},
  \bibinfo {author} {\bibfnamefont {X.}~\bibnamefont {Xu}}, \bibinfo {author}
  {\bibfnamefont {C.~J.}\ \bibnamefont {Fennie}}, \ and\ \bibinfo {author}
  {\bibfnamefont {D.}~\bibnamefont {Xiao}},\ }\bibfield  {title} {\enquote
  {\bibinfo {title} {{Stacking-dependent magnetism in bilayer
  ${\mathrm{CrI}}_{3}$}},}\ }\href {\doibase 10.1021/acs.nanolett.8b03321}
  {\bibfield  {journal} {\bibinfo  {journal} {Nano Lett.}\ }\textbf {\bibinfo
  {volume} {18}},\ \bibinfo {pages} {7658} (\bibinfo {year}
  {2018})}\BibitemShut {NoStop}%
\bibitem [{\citenamefont {Jang}\ \emph {et~al.}(2019)\citenamefont {Jang},
  \citenamefont {Jeong}, \citenamefont {Yoon}, \citenamefont {Ryee},\ and\
  \citenamefont {Han}}]{PhysRevMaterials.3.031001}%
  \BibitemOpen
  \bibfield  {author} {\bibinfo {author} {\bibfnamefont {S.~W.}\ \bibnamefont
  {Jang}}, \bibinfo {author} {\bibfnamefont {M.~Y.}\ \bibnamefont {Jeong}},
  \bibinfo {author} {\bibfnamefont {H.}~\bibnamefont {Yoon}}, \bibinfo {author}
  {\bibfnamefont {S.}~\bibnamefont {Ryee}}, \ and\ \bibinfo {author}
  {\bibfnamefont {M.~J.}\ \bibnamefont {Han}},\ }\bibfield  {title} {\enquote
  {\bibinfo {title} {{Microscopic understanding of magnetic interactions in
  bilayer ${\mathrm{CrI}}_{3}$}},}\ }\href {\doibase
  10.1103/PhysRevMaterials.3.031001} {\bibfield  {journal} {\bibinfo  {journal}
  {Phys. Rev. Materials}\ }\textbf {\bibinfo {volume} {3}},\ \bibinfo {pages}
  {031001} (\bibinfo {year} {2019})}\BibitemShut {NoStop}%
\bibitem [{\citenamefont {Djurdji\ifmmode\acute{c}\else\'{c}\fi{}-Mijin}\ \emph
  {et~al.}(2018)\citenamefont {Djurdji\ifmmode\acute{c}\else\'{c}\fi{}-Mijin},
  \citenamefont {\ifmmode \check{S}\else \v{S}\fi{}olaji\ifmmode~\acute{c}\else
  \'{c}\fi{}}, \citenamefont {Pe\ifmmode \check{s}\else
  \v{s}\fi{}i\ifmmode~\acute{c}\else \'{c}\fi{}}, \citenamefont {\ifmmode
  \check{S}\else \v{S}\fi{}\ifmmode \acute{c}\else
  \'{c}\fi{}epanovi\ifmmode~\acute{c}\else \'{c}\fi{}}, \citenamefont {Liu},
  \citenamefont {Baum}, \citenamefont {Petrovic}, \citenamefont
  {Lazarevi\ifmmode~\acute{c}\else \'{c}\fi{}},\ and\ \citenamefont
  {Popovi\ifmmode~\acute{c}\else \'{c}\fi{}}}]{PhysRevB.98.104307}%
  \BibitemOpen
  \bibfield  {author} {\bibinfo {author} {\bibfnamefont {S.}~\bibnamefont
  {Djurdji\ifmmode\acute{c}\else\'{c}\fi{}-Mijin}}, \bibinfo {author}
  {\bibfnamefont {A.}~\bibnamefont {\ifmmode \check{S}\else
  \v{S}\fi{}olaji\ifmmode~\acute{c}\else \'{c}\fi{}}}, \bibinfo {author}
  {\bibfnamefont {J.}~\bibnamefont {Pe\ifmmode \check{s}\else
  \v{s}\fi{}i\ifmmode~\acute{c}\else \'{c}\fi{}}}, \bibinfo {author}
  {\bibfnamefont {M.}~\bibnamefont {\ifmmode \check{S}\else \v{S}\fi{}\ifmmode
  \acute{c}\else \'{c}\fi{}epanovi\ifmmode~\acute{c}\else \'{c}\fi{}}},
  \bibinfo {author} {\bibfnamefont {Y.}~\bibnamefont {Liu}}, \bibinfo {author}
  {\bibfnamefont {A.}~\bibnamefont {Baum}}, \bibinfo {author} {\bibfnamefont
  {C.}~\bibnamefont {Petrovic}}, \bibinfo {author} {\bibfnamefont
  {N.}~\bibnamefont {Lazarevi\ifmmode~\acute{c}\else \'{c}\fi{}}}, \ and\
  \bibinfo {author} {\bibfnamefont {Z.~V.}\ \bibnamefont
  {Popovi\ifmmode~\acute{c}\else \'{c}\fi{}}},\ }\bibfield  {title} {\enquote
  {\bibinfo {title} {{Lattice dynamics and phase transition in
  ${\mathrm{CrI}}_{3}$ single crystals}},}\ }\href {\doibase
  10.1103/PhysRevB.98.104307} {\bibfield  {journal} {\bibinfo  {journal} {Phys.
  Rev. B}\ }\textbf {\bibinfo {volume} {98}},\ \bibinfo {pages} {104307}
  (\bibinfo {year} {2018})}\BibitemShut {NoStop}%
\bibitem [{\citenamefont {Kim}\ \emph {et~al.}(2018)\citenamefont {Kim},
  \citenamefont {Yang}, \citenamefont {Patel}, \citenamefont {Sfigakis},
  \citenamefont {Li}, \citenamefont {Tian}, \citenamefont {Lei},\ and\
  \citenamefont {Tsen}}]{kim2018one}%
  \BibitemOpen
  \bibfield  {author} {\bibinfo {author} {\bibfnamefont {H.~H.}\ \bibnamefont
  {Kim}}, \bibinfo {author} {\bibfnamefont {B.}~\bibnamefont {Yang}}, \bibinfo
  {author} {\bibfnamefont {T.}~\bibnamefont {Patel}}, \bibinfo {author}
  {\bibfnamefont {F.}~\bibnamefont {Sfigakis}}, \bibinfo {author}
  {\bibfnamefont {C.}~\bibnamefont {Li}}, \bibinfo {author} {\bibfnamefont
  {S.}~\bibnamefont {Tian}}, \bibinfo {author} {\bibfnamefont {H.}~\bibnamefont
  {Lei}}, \ and\ \bibinfo {author} {\bibfnamefont {A.~W.}\ \bibnamefont
  {Tsen}},\ }\bibfield  {title} {\enquote {\bibinfo {title} {{One million
  percent tunnel magnetoresistance in a magnetic van der Waals
  heterostructure}},}\ }\href {\doibase 10.1021/acs.nanolett.8b01552}
  {\bibfield  {journal} {\bibinfo  {journal} {Nano Lett.}\ }\textbf {\bibinfo
  {volume} {18}},\ \bibinfo {pages} {4885} (\bibinfo {year}
  {2018})}\BibitemShut {NoStop}%
\bibitem [{\citenamefont {Jiang}\ \emph
  {et~al.}(2018{\natexlab{a}})\citenamefont {Jiang}, \citenamefont {Li},
  \citenamefont {Wang}, \citenamefont {Mak},\ and\ \citenamefont
  {Shan}}]{ref4}%
  \BibitemOpen
  \bibfield  {author} {\bibinfo {author} {\bibfnamefont {S.}~\bibnamefont
  {Jiang}}, \bibinfo {author} {\bibfnamefont {L.}~\bibnamefont {Li}}, \bibinfo
  {author} {\bibfnamefont {Z.}~\bibnamefont {Wang}}, \bibinfo {author}
  {\bibfnamefont {K.~F.}\ \bibnamefont {Mak}}, \ and\ \bibinfo {author}
  {\bibfnamefont {J.}~\bibnamefont {Shan}},\ }\bibfield  {title} {\enquote
  {\bibinfo {title} {{Controlling magnetism in 2D ${\mathrm{CrI}}_{3}$ by
  electrostatic doping}},}\ }\href {\doibase 10.1038/s41565-018-0135-x}
  {\bibfield  {journal} {\bibinfo  {journal} {Nat. Nanotechnol.}\ }\textbf
  {\bibinfo {volume} {13}},\ \bibinfo {pages} {549} (\bibinfo {year}
  {2018}{\natexlab{a}})}\BibitemShut {NoStop}%
\bibitem [{\citenamefont {Jiang}\ \emph
  {et~al.}(2018{\natexlab{b}})\citenamefont {Jiang}, \citenamefont {Shan},\
  and\ \citenamefont {Mak}}]{ref5}%
  \BibitemOpen
  \bibfield  {author} {\bibinfo {author} {\bibfnamefont {S.}~\bibnamefont
  {Jiang}}, \bibinfo {author} {\bibfnamefont {J.}~\bibnamefont {Shan}}, \ and\
  \bibinfo {author} {\bibfnamefont {K.~F.}\ \bibnamefont {Mak}},\ }\bibfield
  {title} {\enquote {\bibinfo {title} {{Electric-field switching of
  two-dimensional van der Waals magnets}},}\ }\href {\doibase
  10.1038/s41563-018-0040-6} {\bibfield  {journal} {\bibinfo  {journal} {Nat.
  Mater.}\ }\textbf {\bibinfo {volume} {17}},\ \bibinfo {pages} {406} (\bibinfo
  {year} {2018}{\natexlab{b}})}\BibitemShut {NoStop}%
\bibitem [{\citenamefont {Xing}\ \emph {et~al.}(2017)\citenamefont {Xing},
  \citenamefont {Chen}, \citenamefont {Odenthal}, \citenamefont {Zhang},
  \citenamefont {Yuan}, \citenamefont {Su}, \citenamefont {Song}, \citenamefont
  {Wang}, \citenamefont {Zhong}, \citenamefont {Jia} \emph {et~al.}}]{ref10}%
  \BibitemOpen
  \bibfield  {author} {\bibinfo {author} {\bibfnamefont {W.}~\bibnamefont
  {Xing}}, \bibinfo {author} {\bibfnamefont {Y.}~\bibnamefont {Chen}}, \bibinfo
  {author} {\bibfnamefont {P.~M.}\ \bibnamefont {Odenthal}}, \bibinfo {author}
  {\bibfnamefont {X.}~\bibnamefont {Zhang}}, \bibinfo {author} {\bibfnamefont
  {W.}~\bibnamefont {Yuan}}, \bibinfo {author} {\bibfnamefont {T.}~\bibnamefont
  {Su}}, \bibinfo {author} {\bibfnamefont {Q.}~\bibnamefont {Song}}, \bibinfo
  {author} {\bibfnamefont {T.}~\bibnamefont {Wang}}, \bibinfo {author}
  {\bibfnamefont {J.}~\bibnamefont {Zhong}}, \bibinfo {author} {\bibfnamefont
  {S.}~\bibnamefont {Jia}},  \emph {et~al.},\ }\bibfield  {title} {\enquote
  {\bibinfo {title} {{Electric field effect in multilayer Cr$_2$Ge$_2$Te$_6$: a
  ferromagnetic 2D material}},}\ }\href {\doibase 10.1088/2053-1583/aa7034}
  {\bibfield  {journal} {\bibinfo  {journal} {2D Mater.}\ }\textbf {\bibinfo
  {volume} {4}},\ \bibinfo {pages} {024009} (\bibinfo {year}
  {2017})}\BibitemShut {NoStop}%
\bibitem [{\citenamefont {Guo}\ \emph {et~al.}(2018)\citenamefont {Guo},
  \citenamefont {Bi}, \citenamefont {Cai},\ and\ \citenamefont {Wu}}]{ref12}%
  \BibitemOpen
  \bibfield  {author} {\bibinfo {author} {\bibfnamefont {G.}~\bibnamefont
  {Guo}}, \bibinfo {author} {\bibfnamefont {G.}~\bibnamefont {Bi}}, \bibinfo
  {author} {\bibfnamefont {C.}~\bibnamefont {Cai}}, \ and\ \bibinfo {author}
  {\bibfnamefont {H.}~\bibnamefont {Wu}},\ }\bibfield  {title} {\enquote
  {\bibinfo {title} {{Effects of external magnetic field and out-of-plane
  strain on magneto-optical Kerr spectra in ${\mathrm{CrI}}_{3}$ monolayer}},}\
  }\href {\doibase 10.1088/1361-648X/aac96e} {\bibfield  {journal} {\bibinfo
  {journal} {J. Phys. Condens. Matter.}\ }\textbf {\bibinfo {volume} {30}},\
  \bibinfo {pages} {285303} (\bibinfo {year} {2018})}\BibitemShut {NoStop}%
\bibitem [{\citenamefont {Jiang}\ \emph
  {et~al.}(2018{\natexlab{c}})\citenamefont {Jiang}, \citenamefont {Li},
  \citenamefont {Liao}, \citenamefont {Zhao},\ and\ \citenamefont
  {Zhong}}]{ref13}%
  \BibitemOpen
  \bibfield  {author} {\bibinfo {author} {\bibfnamefont {P.}~\bibnamefont
  {Jiang}}, \bibinfo {author} {\bibfnamefont {L.}~\bibnamefont {Li}}, \bibinfo
  {author} {\bibfnamefont {Z.}~\bibnamefont {Liao}}, \bibinfo {author}
  {\bibfnamefont {Y.}~\bibnamefont {Zhao}}, \ and\ \bibinfo {author}
  {\bibfnamefont {Z.}~\bibnamefont {Zhong}},\ }\bibfield  {title} {\enquote
  {\bibinfo {title} {{Spin direction-controlled electronic band structure in
  two-dimensional ferromagnetic ${\mathrm{CrI}}_{3}$}},}\ }\href {\doibase
  10.1021/acs.nanolett.8b01125} {\bibfield  {journal} {\bibinfo  {journal}
  {Nano Lett.}\ }\textbf {\bibinfo {volume} {18}},\ \bibinfo {pages} {3844}
  (\bibinfo {year} {2018}{\natexlab{c}})}\BibitemShut {NoStop}%
\bibitem [{\citenamefont {Zhang}\ \emph {et~al.}(2015)\citenamefont {Zhang},
  \citenamefont {Qu}, \citenamefont {Zhu},\ and\ \citenamefont {Lam}}]{ref14}%
  \BibitemOpen
  \bibfield  {author} {\bibinfo {author} {\bibfnamefont {W.-B.}\ \bibnamefont
  {Zhang}}, \bibinfo {author} {\bibfnamefont {Q.}~\bibnamefont {Qu}}, \bibinfo
  {author} {\bibfnamefont {P.}~\bibnamefont {Zhu}}, \ and\ \bibinfo {author}
  {\bibfnamefont {C.-H.}\ \bibnamefont {Lam}},\ }\bibfield  {title} {\enquote
  {\bibinfo {title} {{Robust intrinsic ferromagnetism and half semiconductivity
  in stable two-dimensional single-layer chromium trihalides}},}\ }\href
  {\doibase 10.1039/C5TC02840J} {\bibfield  {journal} {\bibinfo  {journal} {J.
  Mater. Chem. C}\ }\textbf {\bibinfo {volume} {3}},\ \bibinfo {pages} {12457}
  (\bibinfo {year} {2015})}\BibitemShut {NoStop}%
\bibitem [{\citenamefont {Morell}\ \emph {et~al.}(2019)\citenamefont {Morell},
  \citenamefont {Le{\'o}n}, \citenamefont {Miwa},\ and\ \citenamefont
  {Vargas}}]{ref11}%
  \BibitemOpen
  \bibfield  {author} {\bibinfo {author} {\bibfnamefont {E.~S.}\ \bibnamefont
  {Morell}}, \bibinfo {author} {\bibfnamefont {A.}~\bibnamefont {Le{\'o}n}},
  \bibinfo {author} {\bibfnamefont {R.~H.}\ \bibnamefont {Miwa}}, \ and\
  \bibinfo {author} {\bibfnamefont {P.}~\bibnamefont {Vargas}},\ }\bibfield
  {title} {\enquote {\bibinfo {title} {{Control of magnetism in bilayer
  ${\mathrm{CrI}}_{3}$ by an external electric field}},}\ }\href {\doibase
  10.1088/2053-1583/ab04fb} {\bibfield  {journal} {\bibinfo  {journal} {2D
  Mater.}\ }\textbf {\bibinfo {volume} {6}},\ \bibinfo {pages} {025020}
  (\bibinfo {year} {2019})}\BibitemShut {NoStop}%
\bibitem [{\citenamefont {Liu}\ \emph {et~al.}(2018{\natexlab{a}})\citenamefont
  {Liu}, \citenamefont {Shi}, \citenamefont {Lu},\ and\ \citenamefont
  {Anantram}}]{Liu2018}%
  \BibitemOpen
  \bibfield  {author} {\bibinfo {author} {\bibfnamefont {J.}~\bibnamefont
  {Liu}}, \bibinfo {author} {\bibfnamefont {M.}~\bibnamefont {Shi}}, \bibinfo
  {author} {\bibfnamefont {J.}~\bibnamefont {Lu}}, \ and\ \bibinfo {author}
  {\bibfnamefont {M.~P.}\ \bibnamefont {Anantram}},\ }\bibfield  {title}
  {\enquote {\bibinfo {title} {{Analysis of electrical-field-dependent
  Dzyaloshinskii-Moriya interaction and magnetocrystalline anisotropy in a
  two-dimensional ferromagnetic monolayer}},}\ }\href {\doibase
  10.1103/PhysRevB.97.054416} {\bibfield  {journal} {\bibinfo  {journal} {Phys.
  Rev. B}\ }\textbf {\bibinfo {volume} {97}},\ \bibinfo {pages} {054416}
  (\bibinfo {year} {2018}{\natexlab{a}})}\BibitemShut {NoStop}%
\bibitem [{\citenamefont {Behera}\ \emph {et~al.}(2019)\citenamefont {Behera},
  \citenamefont {Chowdhury},\ and\ \citenamefont {Das}}]{Behera}%
  \BibitemOpen
  \bibfield  {author} {\bibinfo {author} {\bibfnamefont {A.~K.}\ \bibnamefont
  {Behera}}, \bibinfo {author} {\bibfnamefont {S.}~\bibnamefont {Chowdhury}}, \
  and\ \bibinfo {author} {\bibfnamefont {S.~R.}\ \bibnamefont {Das}},\
  }\bibfield  {title} {\enquote {\bibinfo {title} {{Magnetic skyrmions in
  atomic thin ${\mathrm{CrI}}_{3}$ monolayer}},}\ }\href {\doibase
  10.1063/1.5096782} {\bibfield  {journal} {\bibinfo  {journal} {Appl. Phys.
  Lett.}\ }\textbf {\bibinfo {volume} {114}},\ \bibinfo {pages} {232402}
  (\bibinfo {year} {2019})}\BibitemShut {NoStop}%
\bibitem [{\citenamefont {Ghosh}\ \emph {et~al.}(2019)\citenamefont {Ghosh},
  \citenamefont {Stoji{\'c}},\ and\ \citenamefont {Binggeli}}]{Ghosh}%
  \BibitemOpen
  \bibfield  {author} {\bibinfo {author} {\bibfnamefont {S.}~\bibnamefont
  {Ghosh}}, \bibinfo {author} {\bibfnamefont {N.}~\bibnamefont {Stoji{\'c}}}, \
  and\ \bibinfo {author} {\bibfnamefont {N.}~\bibnamefont {Binggeli}},\
  }\bibfield  {title} {\enquote {\bibinfo {title} {{Structural and magnetic
  response of ${\mathrm{CrI}}_{3}$ monolayer to electric field}},}\ }\href
  {\doibase 10.1016/j.physb.2019.06.040} {\bibfield  {journal} {\bibinfo
  {journal} {Physica B Condens. Matter.}\ }\textbf {\bibinfo {volume} {570}},\
  \bibinfo {pages} {166} (\bibinfo {year} {2019})}\BibitemShut {NoStop}%
\bibitem [{\citenamefont {Losada}\ \emph {et~al.}(2019)\citenamefont {Losada},
  \citenamefont {Brataas},\ and\ \citenamefont {Qaiumzadeh}}]{Alireza}%
  \BibitemOpen
  \bibfield  {author} {\bibinfo {author} {\bibfnamefont {J.~M.}\ \bibnamefont
  {Losada}}, \bibinfo {author} {\bibfnamefont {A.}~\bibnamefont {Brataas}}, \
  and\ \bibinfo {author} {\bibfnamefont {A.}~\bibnamefont {Qaiumzadeh}},\
  }\bibfield  {title} {\enquote {\bibinfo {title} {{Ultrafast control of spin
  interactions in honeycomb antiferromagnetic insulators}},}\ }\href {\doibase
  10.1103/PhysRevB.100.060410} {\bibfield  {journal} {\bibinfo  {journal}
  {Phys. Rev. B}\ }\textbf {\bibinfo {volume} {100}},\ \bibinfo {pages}
  {060410} (\bibinfo {year} {2019})}\BibitemShut {NoStop}%
\bibitem [{\citenamefont {Owerre}(2016)}]{owerre2016first}%
  \BibitemOpen
  \bibfield  {author} {\bibinfo {author} {\bibfnamefont {S.}~\bibnamefont
  {Owerre}},\ }\bibfield  {title} {\enquote {\bibinfo {title} {{A first
  theoretical realization of honeycomb topological magnon insulator}},}\ }\href
  {\doibase 10.1088/0953-8984/28/38/386001} {\bibfield  {journal} {\bibinfo
  {journal} {J. Phys. Condens. Matter.}\ }\textbf {\bibinfo {volume} {28}},\
  \bibinfo {pages} {386001} (\bibinfo {year} {2016})}\BibitemShut {NoStop}%
\bibitem [{\citenamefont {Kim}\ \emph {et~al.}(2016)\citenamefont {Kim},
  \citenamefont {Ochoa}, \citenamefont {Zarzuela},\ and\ \citenamefont
  {Tserkovnyak}}]{Yaroslav}%
  \BibitemOpen
  \bibfield  {author} {\bibinfo {author} {\bibfnamefont {S.~K.}\ \bibnamefont
  {Kim}}, \bibinfo {author} {\bibfnamefont {H.}~\bibnamefont {Ochoa}}, \bibinfo
  {author} {\bibfnamefont {R.}~\bibnamefont {Zarzuela}}, \ and\ \bibinfo
  {author} {\bibfnamefont {Y.}~\bibnamefont {Tserkovnyak}},\ }\bibfield
  {title} {\enquote {\bibinfo {title} {{Realization of the Haldane-Kane-Mele
  Model in a System of Localized Spins}},}\ }\href {\doibase
  10.1103/PhysRevLett.117.227201} {\bibfield  {journal} {\bibinfo  {journal}
  {Phys. Rev. Lett.}\ }\textbf {\bibinfo {volume} {117}},\ \bibinfo {pages}
  {227201} (\bibinfo {year} {2016})}\BibitemShut {NoStop}%
\bibitem [{\citenamefont {Leon}\ \emph {et~al.}(2020)\citenamefont {Leon},
  \citenamefont {Gonz{\'a}lez}, \citenamefont {de~Lima}, \citenamefont
  {Mejia},\ and\ \citenamefont {Morell}}]{leon2020strain}%
  \BibitemOpen
  \bibfield  {author} {\bibinfo {author} {\bibfnamefont {A.}~\bibnamefont
  {Leon}}, \bibinfo {author} {\bibfnamefont {J.}~\bibnamefont {Gonz{\'a}lez}},
  \bibinfo {author} {\bibfnamefont {F.~D.~C.}\ \bibnamefont {de~Lima}},
  \bibinfo {author} {\bibfnamefont {J.}~\bibnamefont {Mejia}}, \ and\ \bibinfo
  {author} {\bibfnamefont {E.~S.}\ \bibnamefont {Morell}},\ }\bibfield  {title}
  {\enquote {\bibinfo {title} {{Strain-induced phase transition in
  ${\mathrm{CrI}}_{3}$ bilayers}},}\ }\href {\doibase 10.1088/2053-1583/ab8268}
  {\bibfield  {journal} {\bibinfo  {journal} {2D Mater.}\ }\textbf {\bibinfo
  {volume} {7}},\ \bibinfo {pages} {035008} (\bibinfo {year}
  {2020})}\BibitemShut {NoStop}%
\bibitem [{\citenamefont {Webster}\ and\ \citenamefont
  {Yan}(2018)}]{LucaWebprb}%
  \BibitemOpen
  \bibfield  {author} {\bibinfo {author} {\bibfnamefont {L.}~\bibnamefont
  {Webster}}\ and\ \bibinfo {author} {\bibfnamefont {J.-A.}\ \bibnamefont
  {Yan}},\ }\bibfield  {title} {\enquote {\bibinfo {title} {{Strain-tunable
  magnetic anisotropy in monolayer ${\mathrm{CrCl}}_{3}$,
  ${\mathrm{CrBr}}_{3}$, and ${\mathrm{CrI}}_{3}$}},}\ }\href {\doibase
  10.1103/PhysRevB.98.144411} {\bibfield  {journal} {\bibinfo  {journal} {Phys.
  Rev. B}\ }\textbf {\bibinfo {volume} {98}},\ \bibinfo {pages} {144411}
  (\bibinfo {year} {2018})}\BibitemShut {NoStop}%
\bibitem [{\citenamefont {Liu}\ \emph {et~al.}(2018{\natexlab{b}})\citenamefont
  {Liu}, \citenamefont {Mo}, \citenamefont {Shi}, \citenamefont {Gao},\ and\
  \citenamefont {Lu}}]{liu2018multi}%
  \BibitemOpen
  \bibfield  {author} {\bibinfo {author} {\bibfnamefont {J.}~\bibnamefont
  {Liu}}, \bibinfo {author} {\bibfnamefont {P.}~\bibnamefont {Mo}}, \bibinfo
  {author} {\bibfnamefont {M.}~\bibnamefont {Shi}}, \bibinfo {author}
  {\bibfnamefont {D.}~\bibnamefont {Gao}}, \ and\ \bibinfo {author}
  {\bibfnamefont {J.}~\bibnamefont {Lu}},\ }\bibfield  {title} {\enquote
  {\bibinfo {title} {{Multi-scale analysis of strain-dependent
  magnetocrystalline anisotropy and strain-induced Villari and Nagaoka-Honda
  effects in a two-dimensional ferromagnetic chromium tri-iodide monolayer}},}\
  }\href {\doibase 10.1063/1.5036924} {\bibfield  {journal} {\bibinfo
  {journal} {J. Appl. Phys.}\ }\textbf {\bibinfo {volume} {124}},\ \bibinfo
  {pages} {044303} (\bibinfo {year} {2018}{\natexlab{b}})}\BibitemShut
  {NoStop}%
\bibitem [{\citenamefont {Wu}\ \emph {et~al.}(2019)\citenamefont {Wu},
  \citenamefont {Yu},\ and\ \citenamefont {Yuan}}]{Wu2019}%
  \BibitemOpen
  \bibfield  {author} {\bibinfo {author} {\bibfnamefont {Z.}~\bibnamefont
  {Wu}}, \bibinfo {author} {\bibfnamefont {J.}~\bibnamefont {Yu}}, \ and\
  \bibinfo {author} {\bibfnamefont {S.}~\bibnamefont {Yuan}},\ }\bibfield
  {title} {\enquote {\bibinfo {title} {{Strain-tunable magnetic and electronic
  properties of monolayer ${\mathrm{CrI}}_{3}$}},}\ }\href {\doibase
  10.1039/C8CP07067A} {\bibfield  {journal} {\bibinfo  {journal} {Phys. Chem.
  Chem. Phys.}\ }\textbf {\bibinfo {volume} {21}},\ \bibinfo {pages} {7750}
  (\bibinfo {year} {2019})}\BibitemShut {NoStop}%
\bibitem [{\citenamefont {Larson}\ and\ \citenamefont
  {Kaxiras}(2018)}]{larson2018raman}%
  \BibitemOpen
  \bibfield  {author} {\bibinfo {author} {\bibfnamefont {D.~T.}\ \bibnamefont
  {Larson}}\ and\ \bibinfo {author} {\bibfnamefont {E.}~\bibnamefont
  {Kaxiras}},\ }\bibfield  {title} {\enquote {\bibinfo {title} {{Raman spectrum
  of ${\mathrm{CrI}}_{3}$: An ab initio study}},}\ }\href {\doibase
  10.1103/PhysRevB.98.085406} {\bibfield  {journal} {\bibinfo  {journal} {Phys.
  Rev. B}\ }\textbf {\bibinfo {volume} {98}},\ \bibinfo {pages} {085406}
  (\bibinfo {year} {2018})}\BibitemShut {NoStop}%
\bibitem [{\citenamefont {Giannozzi}\ \emph {et~al.}(2009)\citenamefont
  {Giannozzi}, \citenamefont {Baroni}, \citenamefont {Bonini}, \citenamefont
  {Calandra}, \citenamefont {Car}, \citenamefont {Cavazzoni}, \citenamefont
  {Ceresoli}, \citenamefont {Chiarotti}, \citenamefont {Cococcioni},
  \citenamefont {Dabo} \emph {et~al.}}]{refQE}%
  \BibitemOpen
  \bibfield  {author} {\bibinfo {author} {\bibfnamefont {P.}~\bibnamefont
  {Giannozzi}}, \bibinfo {author} {\bibfnamefont {S.}~\bibnamefont {Baroni}},
  \bibinfo {author} {\bibfnamefont {N.}~\bibnamefont {Bonini}}, \bibinfo
  {author} {\bibfnamefont {M.}~\bibnamefont {Calandra}}, \bibinfo {author}
  {\bibfnamefont {R.}~\bibnamefont {Car}}, \bibinfo {author} {\bibfnamefont
  {C.}~\bibnamefont {Cavazzoni}}, \bibinfo {author} {\bibfnamefont
  {D.}~\bibnamefont {Ceresoli}}, \bibinfo {author} {\bibfnamefont {G.~L.}\
  \bibnamefont {Chiarotti}}, \bibinfo {author} {\bibfnamefont {M.}~\bibnamefont
  {Cococcioni}}, \bibinfo {author} {\bibfnamefont {I.}~\bibnamefont {Dabo}},
  \emph {et~al.},\ }\bibfield  {title} {\enquote {\bibinfo {title} {{QUANTUM
  ESPRESSO: a modular and open-source software project for quantum simulations
  of materials}},}\ }\href {\doibase 10.1088/0953-8984/21/39/395502} {\bibfield
   {journal} {\bibinfo  {journal} {J. Phys. Condens. Matter.}\ }\textbf
  {\bibinfo {volume} {21}},\ \bibinfo {pages} {395502} (\bibinfo {year}
  {2009})}\BibitemShut {NoStop}%
\bibitem [{\citenamefont {Perdew}\ \emph {et~al.}(1996)\citenamefont {Perdew},
  \citenamefont {Burke},\ and\ \citenamefont {Ernzerhof}}]{refpbe}%
  \BibitemOpen
  \bibfield  {author} {\bibinfo {author} {\bibfnamefont {J.~P.}\ \bibnamefont
  {Perdew}}, \bibinfo {author} {\bibfnamefont {K.}~\bibnamefont {Burke}}, \
  and\ \bibinfo {author} {\bibfnamefont {M.}~\bibnamefont {Ernzerhof}},\
  }\bibfield  {title} {\enquote {\bibinfo {title} {{Generalized gradient
  approximation made simple}},}\ }\href {\doibase 10.1103/PhysRevLett.77.3865}
  {\bibfield  {journal} {\bibinfo  {journal} {Phys. Rev. Lett.}\ }\textbf
  {\bibinfo {volume} {77}},\ \bibinfo {pages} {3865} (\bibinfo {year}
  {1996})}\BibitemShut {NoStop}%
\bibitem [{\citenamefont {Lado}\ and\ \citenamefont
  {Fern{\'a}ndez-Rossier}(2017)}]{Hxxz}%
  \BibitemOpen
  \bibfield  {author} {\bibinfo {author} {\bibfnamefont {J.~L.}\ \bibnamefont
  {Lado}}\ and\ \bibinfo {author} {\bibfnamefont {J.}~\bibnamefont
  {Fern{\'a}ndez-Rossier}},\ }\bibfield  {title} {\enquote {\bibinfo {title}
  {{On the origin of magnetic anisotropy in two dimensional
  ${\mathrm{CrI}}_{3}$}},}\ }\href {\doibase 10.1088/2053-1583/aa75ed}
  {\bibfield  {journal} {\bibinfo  {journal} {2D Mater.}\ }\textbf {\bibinfo
  {volume} {4}},\ \bibinfo {pages} {035002} (\bibinfo {year}
  {2017})}\BibitemShut {NoStop}%
\bibitem [{\citenamefont {Chen}\ \emph {et~al.}(2018)\citenamefont {Chen},
  \citenamefont {Chung}, \citenamefont {Gao}, \citenamefont {Chen},
  \citenamefont {Stone}, \citenamefont {Kolesnikov}, \citenamefont {Huang},\
  and\ \citenamefont {Dai}}]{LebingChen}%
  \BibitemOpen
  \bibfield  {author} {\bibinfo {author} {\bibfnamefont {L.}~\bibnamefont
  {Chen}}, \bibinfo {author} {\bibfnamefont {J.-H.}\ \bibnamefont {Chung}},
  \bibinfo {author} {\bibfnamefont {B.}~\bibnamefont {Gao}}, \bibinfo {author}
  {\bibfnamefont {T.}~\bibnamefont {Chen}}, \bibinfo {author} {\bibfnamefont
  {M.~B.}\ \bibnamefont {Stone}}, \bibinfo {author} {\bibfnamefont {A.~I.}\
  \bibnamefont {Kolesnikov}}, \bibinfo {author} {\bibfnamefont
  {Q.}~\bibnamefont {Huang}}, \ and\ \bibinfo {author} {\bibfnamefont
  {P.}~\bibnamefont {Dai}},\ }\bibfield  {title} {\enquote {\bibinfo {title}
  {{Topological Spin Excitations in Honeycomb Ferromagnet
  ${\mathrm{CrI}}_{3}$}},}\ }\href {\doibase 10.1103/PhysRevX.8.041028}
  {\bibfield  {journal} {\bibinfo  {journal} {Phys. Rev. X}\ }\textbf {\bibinfo
  {volume} {8}},\ \bibinfo {pages} {041028} (\bibinfo {year}
  {2018})}\BibitemShut {NoStop}%
\bibitem [{\citenamefont {Elyasi}\ \emph {et~al.}(2019)\citenamefont {Elyasi},
  \citenamefont {Sato},\ and\ \citenamefont {Bauer}}]{elyasi2019topologically}%
  \BibitemOpen
  \bibfield  {author} {\bibinfo {author} {\bibfnamefont {M.}~\bibnamefont
  {Elyasi}}, \bibinfo {author} {\bibfnamefont {K.}~\bibnamefont {Sato}}, \ and\
  \bibinfo {author} {\bibfnamefont {G.~E.}\ \bibnamefont {Bauer}},\ }\bibfield
  {title} {\enquote {\bibinfo {title} {{Topologically nontrivial magnonic
  solitons}},}\ }\href {\doibase 10.1103/PhysRevB.99.134402} {\bibfield
  {journal} {\bibinfo  {journal} {Phys. Rev. B}\ }\textbf {\bibinfo {volume}
  {99}},\ \bibinfo {pages} {134402} (\bibinfo {year} {2019})}\BibitemShut
  {NoStop}%
\bibitem [{\citenamefont {Cheng}\ \emph {et~al.}(2016)\citenamefont {Cheng},
  \citenamefont {Okamoto},\ and\ \citenamefont {Xiao}}]{cheng2016spin}%
  \BibitemOpen
  \bibfield  {author} {\bibinfo {author} {\bibfnamefont {R.}~\bibnamefont
  {Cheng}}, \bibinfo {author} {\bibfnamefont {S.}~\bibnamefont {Okamoto}}, \
  and\ \bibinfo {author} {\bibfnamefont {D.}~\bibnamefont {Xiao}},\ }\bibfield
  {title} {\enquote {\bibinfo {title} {{Spin Nernst effect of magnons in
  collinear antiferromagnets}},}\ }\href {\doibase
  10.1103/PhysRevLett.117.217202} {\bibfield  {journal} {\bibinfo  {journal}
  {Phys. Rev. Lett.}\ }\textbf {\bibinfo {volume} {117}},\ \bibinfo {pages}
  {217202} (\bibinfo {year} {2016})}\BibitemShut {NoStop}%
\bibitem [{\citenamefont {Zyuzin}\ and\ \citenamefont
  {Kovalev}(2016)}]{zyuzin2016magnon}%
  \BibitemOpen
  \bibfield  {author} {\bibinfo {author} {\bibfnamefont {V.~A.}\ \bibnamefont
  {Zyuzin}}\ and\ \bibinfo {author} {\bibfnamefont {A.~A.}\ \bibnamefont
  {Kovalev}},\ }\bibfield  {title} {\enquote {\bibinfo {title} {{Magnon spin
  nernst effect in antiferromagnets}},}\ }\href {\doibase
  10.1103/PhysRevLett.117.217203} {\bibfield  {journal} {\bibinfo  {journal}
  {Phys. Rev. Lett.}\ }\textbf {\bibinfo {volume} {117}},\ \bibinfo {pages}
  {217203} (\bibinfo {year} {2016})}\BibitemShut {NoStop}%
\bibitem [{\citenamefont {Thingstad}\ \emph {et~al.}(2019)\citenamefont
  {Thingstad}, \citenamefont {Kamra}, \citenamefont {Brataas},\ and\
  \citenamefont {Sudb{\o}}}]{thingstad2019chiral}%
  \BibitemOpen
  \bibfield  {author} {\bibinfo {author} {\bibfnamefont {E.}~\bibnamefont
  {Thingstad}}, \bibinfo {author} {\bibfnamefont {A.}~\bibnamefont {Kamra}},
  \bibinfo {author} {\bibfnamefont {A.}~\bibnamefont {Brataas}}, \ and\
  \bibinfo {author} {\bibfnamefont {A.}~\bibnamefont {Sudb{\o}}},\ }\bibfield
  {title} {\enquote {\bibinfo {title} {{Chiral phonon transport induced by
  topological magnons}},}\ }\href {\doibase 10.1103/PhysRevLett.122.107201}
  {\bibfield  {journal} {\bibinfo  {journal} {Phys. Rev. Lett.}\ }\textbf
  {\bibinfo {volume} {122}},\ \bibinfo {pages} {107201} (\bibinfo {year}
  {2019})}\BibitemShut {NoStop}%
\bibitem [{\citenamefont {Li}\ \emph {et~al.}(2014)\citenamefont {Li},
  \citenamefont {Barreteau}, \citenamefont {Castell}, \citenamefont {Silly},\
  and\ \citenamefont {Smogunov}}]{refMAE}%
  \BibitemOpen
  \bibfield  {author} {\bibinfo {author} {\bibfnamefont {D.}~\bibnamefont
  {Li}}, \bibinfo {author} {\bibfnamefont {C.}~\bibnamefont {Barreteau}},
  \bibinfo {author} {\bibfnamefont {M.~R.}\ \bibnamefont {Castell}}, \bibinfo
  {author} {\bibfnamefont {F.}~\bibnamefont {Silly}}, \ and\ \bibinfo {author}
  {\bibfnamefont {A.}~\bibnamefont {Smogunov}},\ }\bibfield  {title} {\enquote
  {\bibinfo {title} {{Out- versus in-plane magnetic anisotropy of free Fe and
  Co nanocrystals: Tight-binding and first-principles studies}},}\ }\href
  {\doibase 10.1103/PhysRevB.90.205409} {\bibfield  {journal} {\bibinfo
  {journal} {Phys. Rev. B}\ }\textbf {\bibinfo {volume} {90}},\ \bibinfo
  {pages} {205409} (\bibinfo {year} {2014})}\BibitemShut {NoStop}%
\bibitem [{\citenamefont {Machintosh}\ and\ \citenamefont
  {Andersen}(1980)}]{andersen19805}%
  \BibitemOpen
  \bibfield  {author} {\bibinfo {author} {\bibfnamefont {A.}~\bibnamefont
  {Machintosh}}\ and\ \bibinfo {author} {\bibfnamefont {O.}~\bibnamefont
  {Andersen}},\ }\href@noop {} {\emph {\bibinfo {title} {{Electrons at the
  Fermi Surface, ed. M. Springford}}}}\ (\bibinfo  {publisher} {Cambridge Univ.
  Press, London},\ \bibinfo {year} {1980})\BibitemShut {NoStop}%
\bibitem [{\citenamefont {Webster}\ \emph {et~al.}(2018)\citenamefont
  {Webster}, \citenamefont {Liang},\ and\ \citenamefont {Yan}}]{refgaptheory}%
  \BibitemOpen
  \bibfield  {author} {\bibinfo {author} {\bibfnamefont {L.}~\bibnamefont
  {Webster}}, \bibinfo {author} {\bibfnamefont {L.}~\bibnamefont {Liang}}, \
  and\ \bibinfo {author} {\bibfnamefont {J.-A.}\ \bibnamefont {Yan}},\
  }\bibfield  {title} {\enquote {\bibinfo {title} {{Distinct spin--lattice and
  spin--phonon interactions in monolayer magnetic ${\mathrm{CrI}}_{3}$}},}\
  }\href {\doibase 10.1039/C8CP03599G} {\bibfield  {journal} {\bibinfo
  {journal} {Phys. Chem. Chem. Phys.}\ }\textbf {\bibinfo {volume} {20}},\
  \bibinfo {pages} {23546} (\bibinfo {year} {2018})}\BibitemShut {NoStop}%
\bibitem [{\citenamefont {Anderson}(1950)}]{anderson1950}%
  \BibitemOpen
  \bibfield  {author} {\bibinfo {author} {\bibfnamefont {P.~W.}\ \bibnamefont
  {Anderson}},\ }\bibfield  {title} {\enquote {\bibinfo {title}
  {{Antiferromagnetism. Theory of Superexchange Interaction}},}\ }\href
  {\doibase 10.1103/PhysRev.79.350} {\bibfield  {journal} {\bibinfo  {journal}
  {Phys. Rev.}\ }\textbf {\bibinfo {volume} {79}},\ \bibinfo {pages} {350}
  (\bibinfo {year} {1950})}\BibitemShut {NoStop}%
\bibitem [{\citenamefont {{Zhang}}\ \emph {et~al.}(2020)\citenamefont
  {{Zhang}}, \citenamefont {{Li}}, \citenamefont {{Weber}}, \citenamefont
  {{Goldberger}}, \citenamefont {{Mak}},\ and\ \citenamefont {{Shan}}}]{Kerr}%
  \BibitemOpen
  \bibfield  {author} {\bibinfo {author} {\bibfnamefont {X.-X.}\ \bibnamefont
  {{Zhang}}}, \bibinfo {author} {\bibfnamefont {L.}~\bibnamefont {{Li}}},
  \bibinfo {author} {\bibfnamefont {D.}~\bibnamefont {{Weber}}}, \bibinfo
  {author} {\bibfnamefont {J.}~\bibnamefont {{Goldberger}}}, \bibinfo {author}
  {\bibfnamefont {K.~F.}\ \bibnamefont {{Mak}}}, \ and\ \bibinfo {author}
  {\bibfnamefont {J.}~\bibnamefont {{Shan}}},\ }\bibfield  {title} {\enquote
  {\bibinfo {title} {{Gate-tunable spin waves in antiferromagnetic atomic
  bilayers}},}\ }\href@noop {} {\bibfield  {journal} {\bibinfo  {journal}
  {arXiv e-prints}\ ,\ \bibinfo {eid} {arXiv:2001.04044}} (\bibinfo {year}
  {2020})},\ \Eprint {http://arxiv.org/abs/2001.04044} {arXiv:2001.04044
  [cond-mat.mes-hall]} \BibitemShut {NoStop}%
\bibitem [{\citenamefont {{Cenker}}\ \emph {et~al.}(2020)\citenamefont
  {{Cenker}}, \citenamefont {{Huang}}, \citenamefont {{Suri}}, \citenamefont
  {{Thijssen}}, \citenamefont {{Miller}}, \citenamefont {{Song}}, \citenamefont
  {{Taniguchi}}, \citenamefont {{Watanabe}}, \citenamefont {{McGuire}},
  \citenamefont {{Xiao}},\ and\ \citenamefont {{Xu}}}]{Raman}%
  \BibitemOpen
  \bibfield  {author} {\bibinfo {author} {\bibfnamefont {J.}~\bibnamefont
  {{Cenker}}}, \bibinfo {author} {\bibfnamefont {B.}~\bibnamefont {{Huang}}},
  \bibinfo {author} {\bibfnamefont {N.}~\bibnamefont {{Suri}}}, \bibinfo
  {author} {\bibfnamefont {P.}~\bibnamefont {{Thijssen}}}, \bibinfo {author}
  {\bibfnamefont {A.}~\bibnamefont {{Miller}}}, \bibinfo {author}
  {\bibfnamefont {T.}~\bibnamefont {{Song}}}, \bibinfo {author} {\bibfnamefont
  {T.}~\bibnamefont {{Taniguchi}}}, \bibinfo {author} {\bibfnamefont
  {K.}~\bibnamefont {{Watanabe}}}, \bibinfo {author} {\bibfnamefont {M.~A.}\
  \bibnamefont {{McGuire}}}, \bibinfo {author} {\bibfnamefont {D.}~\bibnamefont
  {{Xiao}}}, \ and\ \bibinfo {author} {\bibfnamefont {X.}~\bibnamefont
  {{Xu}}},\ }\bibfield  {title} {\enquote {\bibinfo {title} {{Direct
  observation of 2D magnons in atomically thin ${\mathrm{CrI}}_{3}$}},}\
  }\href@noop {} {\bibfield  {journal} {\bibinfo  {journal} {arXiv e-prints}\
  ,\ \bibinfo {eid} {arXiv:2001.07025}} (\bibinfo {year} {2020})},\ \Eprint
  {http://arxiv.org/abs/2001.07025} {arXiv:2001.07025 [cond-mat.mes-hall]}
  \BibitemShut {NoStop}%
\bibitem [{\citenamefont {Nembach}\ \emph {et~al.}(2015)\citenamefont
  {Nembach}, \citenamefont {Shaw}, \citenamefont {Weiler}, \citenamefont
  {Ju{\'e}},\ and\ \citenamefont {Silva}}]{nembach2015linear}%
  \BibitemOpen
  \bibfield  {author} {\bibinfo {author} {\bibfnamefont {H.~T.}\ \bibnamefont
  {Nembach}}, \bibinfo {author} {\bibfnamefont {J.~M.}\ \bibnamefont {Shaw}},
  \bibinfo {author} {\bibfnamefont {M.}~\bibnamefont {Weiler}}, \bibinfo
  {author} {\bibfnamefont {E.}~\bibnamefont {Ju{\'e}}}, \ and\ \bibinfo
  {author} {\bibfnamefont {T.~J.}\ \bibnamefont {Silva}},\ }\bibfield  {title}
  {\enquote {\bibinfo {title} {{Linear relation between Heisenberg exchange and
  interfacial Dzyaloshinskii--Moriya interaction in metal films}},}\ }\href
  {\doibase 10.1038/nphys3418} {\bibfield  {journal} {\bibinfo  {journal} {Nat.
  Phys.}\ }\textbf {\bibinfo {volume} {11}},\ \bibinfo {pages} {825} (\bibinfo
  {year} {2015})}\BibitemShut {NoStop}%
\bibitem [{\citenamefont {Di}\ \emph {et~al.}(2015)\citenamefont {Di},
  \citenamefont {Zhang}, \citenamefont {Lim}, \citenamefont {Ng}, \citenamefont
  {Kuok}, \citenamefont {Yu}, \citenamefont {Yoon}, \citenamefont {Qiu},\ and\
  \citenamefont {Yang}}]{di2015direct}%
  \BibitemOpen
  \bibfield  {author} {\bibinfo {author} {\bibfnamefont {K.}~\bibnamefont
  {Di}}, \bibinfo {author} {\bibfnamefont {V.~L.}\ \bibnamefont {Zhang}},
  \bibinfo {author} {\bibfnamefont {H.~S.}\ \bibnamefont {Lim}}, \bibinfo
  {author} {\bibfnamefont {S.~C.}\ \bibnamefont {Ng}}, \bibinfo {author}
  {\bibfnamefont {M.~H.}\ \bibnamefont {Kuok}}, \bibinfo {author}
  {\bibfnamefont {J.}~\bibnamefont {Yu}}, \bibinfo {author} {\bibfnamefont
  {J.}~\bibnamefont {Yoon}}, \bibinfo {author} {\bibfnamefont {X.}~\bibnamefont
  {Qiu}}, \ and\ \bibinfo {author} {\bibfnamefont {H.}~\bibnamefont {Yang}},\
  }\bibfield  {title} {\enquote {\bibinfo {title} {{Direct observation of the
  Dzyaloshinskii-Moriya interaction in a Pt/Co/Ni film}},}\ }\href {\doibase
  10.1103/PhysRevLett.114.047201} {\bibfield  {journal} {\bibinfo  {journal}
  {Phys. Rev. Lett.}\ }\textbf {\bibinfo {volume} {114}},\ \bibinfo {pages}
  {047201} (\bibinfo {year} {2015})}\BibitemShut {NoStop}%
\bibitem [{\citenamefont {Belmeguenai}\ \emph {et~al.}(2015)\citenamefont
  {Belmeguenai}, \citenamefont {Adam}, \citenamefont {Roussign{\'e}},
  \citenamefont {Eimer}, \citenamefont {Devolder}, \citenamefont {Kim},
  \citenamefont {Cherif}, \citenamefont {Stashkevich},\ and\ \citenamefont
  {Thiaville}}]{belmeguenai2015interfacial}%
  \BibitemOpen
  \bibfield  {author} {\bibinfo {author} {\bibfnamefont {M.}~\bibnamefont
  {Belmeguenai}}, \bibinfo {author} {\bibfnamefont {J.-P.}\ \bibnamefont
  {Adam}}, \bibinfo {author} {\bibfnamefont {Y.}~\bibnamefont {Roussign{\'e}}},
  \bibinfo {author} {\bibfnamefont {S.}~\bibnamefont {Eimer}}, \bibinfo
  {author} {\bibfnamefont {T.}~\bibnamefont {Devolder}}, \bibinfo {author}
  {\bibfnamefont {J.-V.}\ \bibnamefont {Kim}}, \bibinfo {author} {\bibfnamefont
  {S.~M.}\ \bibnamefont {Cherif}}, \bibinfo {author} {\bibfnamefont
  {A.}~\bibnamefont {Stashkevich}}, \ and\ \bibinfo {author} {\bibfnamefont
  {A.}~\bibnamefont {Thiaville}},\ }\bibfield  {title} {\enquote {\bibinfo
  {title} {{Interfacial Dzyaloshinskii-Moriya interaction in perpendicularly
  magnetized Pt/Co/AlO$_x$ ultrathin films measured by Brillouin light
  spectroscopy}},}\ }\href {\doibase 10.1103/PhysRevB.91.180405} {\bibfield
  {journal} {\bibinfo  {journal} {Phys. Rev. B}\ }\textbf {\bibinfo {volume}
  {91}},\ \bibinfo {pages} {180405} (\bibinfo {year} {2015})}\BibitemShut
  {NoStop}%
\bibitem [{\citenamefont {Chen}\ \emph {et~al.}(2020)\citenamefont {Chen},
  \citenamefont {Chung}, \citenamefont {Chen}, \citenamefont {Duan},
  \citenamefont {Schneidewind}, \citenamefont {Radelytskyi}, \citenamefont
  {Voneshen}, \citenamefont {Ewings}, \citenamefont {Stone}, \citenamefont
  {Kolesnikov} \emph {et~al.}}]{chen2020magnetic}%
  \BibitemOpen
  \bibfield  {author} {\bibinfo {author} {\bibfnamefont {L.}~\bibnamefont
  {Chen}}, \bibinfo {author} {\bibfnamefont {J.-H.}\ \bibnamefont {Chung}},
  \bibinfo {author} {\bibfnamefont {T.}~\bibnamefont {Chen}}, \bibinfo {author}
  {\bibfnamefont {C.}~\bibnamefont {Duan}}, \bibinfo {author} {\bibfnamefont
  {A.}~\bibnamefont {Schneidewind}}, \bibinfo {author} {\bibfnamefont
  {I.}~\bibnamefont {Radelytskyi}}, \bibinfo {author} {\bibfnamefont {D.~J.}\
  \bibnamefont {Voneshen}}, \bibinfo {author} {\bibfnamefont {R.~A.}\
  \bibnamefont {Ewings}}, \bibinfo {author} {\bibfnamefont {M.~B.}\
  \bibnamefont {Stone}}, \bibinfo {author} {\bibfnamefont {A.~I.}\ \bibnamefont
  {Kolesnikov}},  \emph {et~al.},\ }\bibfield  {title} {\enquote {\bibinfo
  {title} {{Magnetic anisotropy in ferromagnetic ${\mathrm{CrI}}_{3}$}},}\
  }\href {\doibase 10.1103/PhysRevB.101.134418} {\bibfield  {journal} {\bibinfo
   {journal} {Phys. Rev. B}\ }\textbf {\bibinfo {volume} {101}},\ \bibinfo
  {pages} {134418} (\bibinfo {year} {2020})}\BibitemShut {NoStop}%
\bibitem [{\citenamefont {Gitgeatpong}\ \emph {et~al.}(2017)\citenamefont
  {Gitgeatpong}, \citenamefont {Zhao}, \citenamefont {Piyawongwatthana},
  \citenamefont {Qiu}, \citenamefont {Harriger}, \citenamefont {Butch},
  \citenamefont {Sato},\ and\ \citenamefont {Matan}}]{INS-AFM}%
  \BibitemOpen
  \bibfield  {author} {\bibinfo {author} {\bibfnamefont {G.}~\bibnamefont
  {Gitgeatpong}}, \bibinfo {author} {\bibfnamefont {Y.}~\bibnamefont {Zhao}},
  \bibinfo {author} {\bibfnamefont {P.}~\bibnamefont {Piyawongwatthana}},
  \bibinfo {author} {\bibfnamefont {Y.}~\bibnamefont {Qiu}}, \bibinfo {author}
  {\bibfnamefont {L.~W.}\ \bibnamefont {Harriger}}, \bibinfo {author}
  {\bibfnamefont {N.~P.}\ \bibnamefont {Butch}}, \bibinfo {author}
  {\bibfnamefont {T.~J.}\ \bibnamefont {Sato}}, \ and\ \bibinfo {author}
  {\bibfnamefont {K.}~\bibnamefont {Matan}},\ }\bibfield  {title} {\enquote
  {\bibinfo {title} {{Nonreciprocal Magnons and Symmetry-Breaking in the
  Noncentrosymmetric Antiferromagnet}},}\ }\href {\doibase
  10.1103/PhysRevLett.119.047201} {\bibfield  {journal} {\bibinfo  {journal}
  {Phys. Rev. Lett.}\ }\textbf {\bibinfo {volume} {119}},\ \bibinfo {pages}
  {047201} (\bibinfo {year} {2017})}\BibitemShut {NoStop}%
\bibitem [{\citenamefont {Mena}\ \emph {et~al.}(2014)\citenamefont {Mena},
  \citenamefont {Perry}, \citenamefont {Perring}, \citenamefont {Le},
  \citenamefont {Guerrero}, \citenamefont {Storni}, \citenamefont {Adroja},
  \citenamefont {R{\"u}egg},\ and\ \citenamefont {McMorrow}}]{mena2014spin}%
  \BibitemOpen
  \bibfield  {author} {\bibinfo {author} {\bibfnamefont {M.}~\bibnamefont
  {Mena}}, \bibinfo {author} {\bibfnamefont {R.}~\bibnamefont {Perry}},
  \bibinfo {author} {\bibfnamefont {T.}~\bibnamefont {Perring}}, \bibinfo
  {author} {\bibfnamefont {M.}~\bibnamefont {Le}}, \bibinfo {author}
  {\bibfnamefont {S.}~\bibnamefont {Guerrero}}, \bibinfo {author}
  {\bibfnamefont {M.}~\bibnamefont {Storni}}, \bibinfo {author} {\bibfnamefont
  {D.}~\bibnamefont {Adroja}}, \bibinfo {author} {\bibfnamefont
  {C.}~\bibnamefont {R{\"u}egg}}, \ and\ \bibinfo {author} {\bibfnamefont
  {D.}~\bibnamefont {McMorrow}},\ }\bibfield  {title} {\enquote {\bibinfo
  {title} {{Spin-wave spectrum of the quantum ferromagnet on the pyrochlore
  lattice ${\mathrm{Lu}}_{2}{\mathrm{V}}_{2}{\mathrm{O}}_{7}$}},}\ }\href
  {\doibase 10.1103/PhysRevLett.113.047202} {\bibfield  {journal} {\bibinfo
  {journal} {Phys. Rev. Lett.}\ }\textbf {\bibinfo {volume} {113}},\ \bibinfo
  {pages} {047202} (\bibinfo {year} {2014})}\BibitemShut {NoStop}%
\bibitem [{\citenamefont {Flovik}\ \emph {et~al.}(2017)\citenamefont {Flovik},
  \citenamefont {Qaiumzadeh}, \citenamefont {Nandy}, \citenamefont {Heo},\ and\
  \citenamefont {Rasing}}]{skyrmion1}%
  \BibitemOpen
  \bibfield  {author} {\bibinfo {author} {\bibfnamefont {V.}~\bibnamefont
  {Flovik}}, \bibinfo {author} {\bibfnamefont {A.}~\bibnamefont {Qaiumzadeh}},
  \bibinfo {author} {\bibfnamefont {A.~K.}\ \bibnamefont {Nandy}}, \bibinfo
  {author} {\bibfnamefont {C.}~\bibnamefont {Heo}}, \ and\ \bibinfo {author}
  {\bibfnamefont {T.}~\bibnamefont {Rasing}},\ }\bibfield  {title} {\enquote
  {\bibinfo {title} {{Generation of single skyrmions by picosecond magnetic
  field pulses}},}\ }\href {\doibase 10.1103/PhysRevB.96.140411} {\bibfield
  {journal} {\bibinfo  {journal} {Phys. Rev. B}\ }\textbf {\bibinfo {volume}
  {96}},\ \bibinfo {pages} {140411} (\bibinfo {year} {2017})}\BibitemShut
  {NoStop}%
\bibitem [{\citenamefont {Khoshlahni}\ \emph {et~al.}(2019)\citenamefont
  {Khoshlahni}, \citenamefont {Qaiumzadeh}, \citenamefont {Bergman},\ and\
  \citenamefont {Brataas}}]{skyrmion2}%
  \BibitemOpen
  \bibfield  {author} {\bibinfo {author} {\bibfnamefont {R.}~\bibnamefont
  {Khoshlahni}}, \bibinfo {author} {\bibfnamefont {A.}~\bibnamefont
  {Qaiumzadeh}}, \bibinfo {author} {\bibfnamefont {A.}~\bibnamefont {Bergman}},
  \ and\ \bibinfo {author} {\bibfnamefont {A.}~\bibnamefont {Brataas}},\
  }\bibfield  {title} {\enquote {\bibinfo {title} {{Ultrafast generation and
  dynamics of isolated skyrmions in antiferromagnetic insulators}},}\ }\href
  {\doibase 10.1103/PhysRevB.99.054423} {\bibfield  {journal} {\bibinfo
  {journal} {Phys. Rev. B}\ }\textbf {\bibinfo {volume} {99}},\ \bibinfo
  {pages} {054423} (\bibinfo {year} {2019})}\BibitemShut {NoStop}%
\bibitem [{\citenamefont {Thiaville}\ \emph {et~al.}(2012)\citenamefont
  {Thiaville}, \citenamefont {Rohart}, \citenamefont {Ju{\'e}}, \citenamefont
  {Cros},\ and\ \citenamefont {Fert}}]{thiaville2012dynamics}%
  \BibitemOpen
  \bibfield  {author} {\bibinfo {author} {\bibfnamefont {A.}~\bibnamefont
  {Thiaville}}, \bibinfo {author} {\bibfnamefont {S.}~\bibnamefont {Rohart}},
  \bibinfo {author} {\bibfnamefont {{\'E}.}~\bibnamefont {Ju{\'e}}}, \bibinfo
  {author} {\bibfnamefont {V.}~\bibnamefont {Cros}}, \ and\ \bibinfo {author}
  {\bibfnamefont {A.}~\bibnamefont {Fert}},\ }\bibfield  {title} {\enquote
  {\bibinfo {title} {{Dynamics of Dzyaloshinskii domain walls in ultrathin
  magnetic films}},}\ }\href {\doibase 10.1209/0295-5075/100/57002} {\bibfield
  {journal} {\bibinfo  {journal} {EPL (Europhysics Letters)}\ }\textbf
  {\bibinfo {volume} {100}},\ \bibinfo {pages} {57002} (\bibinfo {year}
  {2012})}\BibitemShut {NoStop}%
\bibitem [{\citenamefont {Qaiumzadeh}\ \emph {et~al.}(2018)\citenamefont
  {Qaiumzadeh}, \citenamefont {Kristiansen},\ and\ \citenamefont
  {Brataas}}]{qaiumzadeh2018controlling}%
  \BibitemOpen
  \bibfield  {author} {\bibinfo {author} {\bibfnamefont {A.}~\bibnamefont
  {Qaiumzadeh}}, \bibinfo {author} {\bibfnamefont {L.~A.}\ \bibnamefont
  {Kristiansen}}, \ and\ \bibinfo {author} {\bibfnamefont {A.}~\bibnamefont
  {Brataas}},\ }\bibfield  {title} {\enquote {\bibinfo {title} {{Controlling
  chiral domain walls in antiferromagnets using spin-wave helicity}},}\ }\href
  {\doibase 10.1103/PhysRevB.97.020402} {\bibfield  {journal} {\bibinfo
  {journal} {Phys. Rev. B}\ }\textbf {\bibinfo {volume} {97}},\ \bibinfo
  {pages} {020402} (\bibinfo {year} {2018})}\BibitemShut {NoStop}%
\bibitem [{\citenamefont {Ryu}\ \emph {et~al.}(2013)\citenamefont {Ryu},
  \citenamefont {Thomas}, \citenamefont {Yang},\ and\ \citenamefont
  {Parkin}}]{ryu2013chiral}%
  \BibitemOpen
  \bibfield  {author} {\bibinfo {author} {\bibfnamefont {K.-S.}\ \bibnamefont
  {Ryu}}, \bibinfo {author} {\bibfnamefont {L.}~\bibnamefont {Thomas}},
  \bibinfo {author} {\bibfnamefont {S.-H.}\ \bibnamefont {Yang}}, \ and\
  \bibinfo {author} {\bibfnamefont {S.}~\bibnamefont {Parkin}},\ }\bibfield
  {title} {\enquote {\bibinfo {title} {Chiral spin torque at magnetic domain
  walls},}\ }\href {\doibase 10.1038/nnano.2013.102} {\bibfield  {journal}
  {\bibinfo  {journal} {Nat. Nanotechnol.}\ }\textbf {\bibinfo {volume} {8}},\
  \bibinfo {pages} {527} (\bibinfo {year} {2013})}\BibitemShut {NoStop}%
\end{thebibliography}%

\end{document}